\documentclass[preprint,twocolumn]{aastex63}
\usepackage{float, bm, graphicx, amsmath, morefloats}
\usepackage[caption=false]{subfig}
\shorttitle{Ca-rich Type IIb SNe}
\shortauthors{Das et al.}

\usepackage{float,graphicx,amsmath,tabularx,booktabs,natbib,threeparttable}
\usepackage{float, bm, graphicx, amsmath, morefloats}
\usepackage[caption=false]{subfig}

\usepackage{comment}
\usepackage{lineno}

\usepackage{placeins}

\definecolor{dark-red}{rgb}{0.4,0.15,0.15}
\definecolor{dark-blue}{rgb}{0.15,0.15,0.4}
\definecolor{medium-blue}{rgb}{0,0,0.5}
\hypersetup{
    colorlinks, linkcolor={dark-red},
    citecolor={dark-blue}, urlcolor={medium-blue}
}

\newcommand{\beqa}{\begin{eqnarray}} 
\newcommand{\eeqa}{\end{eqnarray}}

\newcommand{\bsub}{\begin{subequations}}
\newcommand{\esub}{\end{subequations}}
\newcommand{\beal}{\begin{align}}
\newcommand{\ealn}{\end{align}}
\newcommand{\Nif}{$\rm ^{56}Ni$} 
\newcommand{\Cif}{$\rm ^{56}Co$}

\newcommand{\msun}{M$_{\sun}$}

\newcommand{\Mni}{$M_\mathrm{Ni}$}
\newcommand{\Mej}{$M_\mathrm{ej}$}

\newcommand{\cigale}{\textsc{cigale}}
\newcommand{\emcee}{\textsc{emcee}}

\newcommand{\fornax}{\textsc{fornax}}

\newcommand{\lpipe}{\textsc{lpipe}}

\newcommand{\Ni}{\ensuremath{^{56}\mathrm{Ni}}}
\newcommand{\Msun}{{\ensuremath{\mathrm{M}_{\odot}}}}
\newcommand{\Rsun}{{\ensuremath{\mathrm{R}_{\odot}}}}
\newcommand{\galex}{\textit{GALEX}}
\newcommand{\wise}{\textit{WISE}}
\newcommand{\HEAsoft}{\textsc{HEAsoft}}

\graphicspath{{./}{figures/}}

\begin{document}

\title{Probing the low-mass end of core-collapse supernovae using a sample of strongly-stripped Calcium-rich Type IIb Supernovae from the Zwicky Transient Facility}

\author[0000-0001-8372-997X]{Kaustav K.~Das}
\affiliation{Cahill Center for Astrophysics, California Institute of Technology, MC 249-17, 
1200 E California Boulevard, Pasadena, CA, 91125, USA}

\author[0000-0002-5619-4938]{Mansi~M.~Kasliwal}
\affiliation{Cahill Center for Astrophysics, 
California Institute of Technology, MC 249-17, 
1200 E California Boulevard, Pasadena, CA, 91125, USA}

\author[0000-0002-4223-103X]{Christoffer Fremling}
\affiliation{Cahill Center for Astrophysics, 
California Institute of Technology, MC 249-17, 
1200 E California Boulevard, Pasadena, CA, 91125, USA}

\author[0000-0003-1546-6615]{Sheng Yang}
\affiliation{The Oskar Klein Centre, Department of Astronomy, Stockholm University, AlbaNova, SE-10691 Stockholm, Sweden}

\author[0000-0001-6797-1889]{Steve Schulze}
\affiliation{The Oskar Klein Centre, Department of Physics, Stockholm University, AlbaNova, SE-10691 Stockholm, Sweden}

\author[0000-0003-1546-6615]{Jesper Sollerman}
\affiliation{The Oskar Klein Centre, Department of Astronomy, Stockholm University, AlbaNova, SE-10691 Stockholm, Sweden}

\author{Tawny Sit}
\affiliation{Cahill Center for Astrophysics, 
California Institute of Technology, MC 249-17, 
1200 E California Boulevard, Pasadena, CA, 91125, USA}

\author[0000-0002-8989-0542]{Kishalay De}
\affiliation{MIT-Kavli Institute for Astrophysics and Space Research
77 Massachusetts Ave. Cambridge, MA 02139, USA}

\author{Anastasios Tzanidakis}
\affiliation{Cahill Center for Astrophysics, 
California Institute of Technology, MC 249-17, 
1200 E California Boulevard, Pasadena, CA, 91125, USA}

\author[0000-0001-8472-1996]{Daniel A.~Perley}
\affiliation{Astrophysics Research Institute, Liverpool John Moores University, IC2, Liverpool Science Park, Liverpool L3 5RF, UK}


\author{Shreya Anand}
\affiliation{Division of Physics, Mathematics and Astronomy, California Institute of Technology, Pasadena, CA 91125, USA}

\author{Igor Andreoni}
\affiliation{Joint Space-Science Institute, University of Maryland, College Park, MD 20742, USA}
\affiliation{Department of Astronomy, University of Maryland, College Park, MD 20742, USA}
\affiliation{Astrophysics Science Division, NASA Goddard Space Flight Center, Mail Code 661, Greenbelt, MD 20771, USA}

\author{C. Barbarino}
\affiliation{The Oskar Klein Centre, Department of Astronomy, Stockholm University, AlbaNova, SE-10691 Stockholm, Sweden}

\author{K. Brudge}

\author{Andrew Drake}
\affiliation{Cahill Center for Astrophysics, 
California Institute of Technology, MC 249-17, 
1200 E California Boulevard, Pasadena, CA, 91125, USA}

\author{Avishay Gal-Yam}
\affiliation{Department of Particle Physics and Astrophysics, Weizmann Institute of Science, 234 Herzl St, 76100 Rehovot, Israel}

\author[0000-0003-2451-5482]{Russ R. Laher}
\affiliation{IPAC, California Institute of Technology, 1200 E. California
             Blvd, Pasadena, CA 91125, USA}

\author{Viraj~Karambelkar}
\affiliation{Cahill Center for Astrophysics, 
California Institute of Technology, MC 249-17, 
1200 E California Boulevard, Pasadena, CA, 91125, USA}

\author[0000-0001-5390-8563]{S.~R.~Kulkarni}
\affiliation{Cahill Center for Astrophysics, 
California Institute of Technology, MC 249-17, 
1200 E California Boulevard, Pasadena, CA, 91125, USA}

\author[0000-0002-8532-9395]{Frank J. Masci}
\affiliation{IPAC, California Institute of Technology, 1200 E. California
             Blvd, Pasadena, CA 91125, USA}

\author[0000-0002-7226-0659]{Michael S. Medford}
\affiliation{Department of Astronomy, University of California, Berkeley, Berkeley, CA 94720}
\affiliation{Lawrence Berkeley National Laboratory, 1 Cyclotron Rd., Berkeley, CA 94720}

\author{Abigail Polin}
\affiliation{Cahill Center for Astrophysics, 
California Institute of Technology, MC 249-17, 
1200 E California Boulevard, Pasadena, CA, 91125, USA}

\author{Harrison Reedy}
\affiliation{Cahill Center for Astrophysics, 
California Institute of Technology, MC 249-17, 
1200 E California Boulevard, Pasadena, CA, 91125, USA}

\author{Reed Riddle}
\affiliation{Cahill Center for Astrophysics, 
California Institute of Technology, MC 249-17, 
1200 E California Boulevard, Pasadena, CA, 91125, USA}

\author{Yashvi Sharma}
\affiliation{Cahill Center for Astrophysics, 
California Institute of Technology, MC 249-17, 
1200 E California Boulevard, Pasadena, CA, 91125, USA}

\author[0000-0001-7062-9726]{Roger Smith}
\affiliation{Caltech Optical Observatories, California Institute of Technology, Pasadena, CA 91125, USA}

\author[0000-0003-1710-9339]{Lin~Yan}
\affil{Caltech Optical Observatories, California Institute of Technology, Pasadena, CA 91125, USA}

\author{Yi Yang}

\author[0000-0001-6747-8509]{Yuhan Yao}
\affiliation{Cahill Center for Astrophysics, 
California Institute of Technology, MC 249-17, 
1200 E California Boulevard, Pasadena, CA, 91125, USA}


\correspondingauthor{Kaustav K. Das}
\email{kdas@astro.caltech.edu}


\begin{abstract}
The fate of stars in the zero-age main-sequence (ZAMS) range $\approx 8-12$ \Msun\ is unclear. They could evolve to form white dwarfs or explode as electron-capture supernovae (SNe) or iron core-collapse SNe (CCSNe). Even though the initial mass function indicates that this mass range should account for over 40\% of all CCSNe progenitors, few have been observationally confirmed, likely owing to the faintness and rapid evolution of some of these transients. In this paper, we present a sample of nine Ca-rich/O-poor Type IIb SNe detected by the Zwicky Transient Facility with progenitors likely in this mass range. These sources have a flux ratio of [Ca II] $\lambda \lambda$7291, 7324 to [O I] $\lambda \lambda$6300, 6364 of $\gtrsim$ 2 in their nebular spectra. Comparing the measured [O I] luminosity ($\lesssim 10^{39} \mathrm{erg\ s^{-1}}$) and derived oxygen mass ($\approx 0.01$ \Msun) with theoretical models, we infer that the progenitor ZAMS mass for these explosions is less than 12 \msun. The ejecta properties (\Mej $\lesssim 1\ \Msun$ and $\mathrm{E_{kin}} \sim 10^{50}$ erg) are also consistent. The low ejecta mass of these sources indicates a class of strongly-stripped SNe that is a transition between the regular stripped-envelope SNe and ultra-stripped SNe. The progenitor could be stripped by a main sequence companion and result in the formation of a neutron star $-$ main sequence binary. Such binaries have been suggested to be progenitors of neutron star $-$ white dwarf systems that could merge within a Hubble time, and be detectable with LISA.

\end{abstract}


\section{Introduction} \label{sec:intro}

The fate of stars with zero-age main-sequence (ZAMS) masses larger than $\approx$ 12 \msun\ is reasonably well understood from stellar evolution theory and observations. They form a massive iron core and may explode as a core-collapse supernova (CCSN) once their cores reach the Chandrasekhar mass. However, the fate of less massive stars in the ZAMS range $\approx 8-12$ \msun\ is less clear \citep[e.g.,][]{Janka2012}. They might form white dwarfs (WD), explode as electron capture supernovae (ECSNe) with an ONeMg core \citep[e.g.,][]{Nomoto1984, Stockinger2020},  or form an iron core like regular massive stars. Depending on the metallicity and mass loss in binary systems, these stars might undergo violent flashes and power unusual transients before their deaths \citep{Woosley2015, Pumo2009, Podsiadlowski2004}.  

We expect that the narrow range of ZAMS masses from $\approx 8-12$ \msun\ should account for a significant fraction of CCSNe, given that the initial mass function (IMF) of stars drops rapidly toward high masses. For a typical IMF, we expect over 40\% of the potential CCSN progenitors to reside in this mass range \citep{Sukhbold2016}. However, surprisingly, the number of SNe associated with a low-mass progenitor is small, with most of them being low-luminosity Type II SNe \citep[e.g.,][]{Sheng2021, Oneill2021, Reguitti2021, Valerin2022, Kozyreva2022}. Despite higher abundance in the universe, transients with low-mass progenitors may be missed because they are faint and evolve rapidly. The low mass end of the CCSN population is expected to be under-energetic and possibly underluminous \citep{Melson2015, Nomoto1982}. Thus, there is likely an observational bias due to the challenge to discover and follow-up of such faint and rapidly-evolving transients. It is difficult to understand the explosion mechanisms and trends for this population due to the small number of detections.

In this paper, we present a sample of Ca-rich Type IIb SNe, with potential progenitor mass lying in this low mass range, detected by the Zwicky Transient Facility \citep[ZTF;][]{Bellm2019, Graham2019}. Type IIb is a class of CCSNe whose progenitors initially exhibit hydrogen spectral features that weaken and disappear in the weeks following explosion. Unlike Type II SNe, which are associated with red-supergiant progenitors with ZAMS mass ranging from 8-20 \Msun \citep[see][for a review]{Smartt:2009, Smartt15, Dyk2017}, the mechanisms that drive the stripping and the regimes in which they dominate are still open questions for Type IIb SNe \citep[e.g.,][]{Sravan2019}.

Ca-rich transients are a class of supernovae that are identified by a high [Ca~II]/[O I] flux ratio ($>$ 2)  in the nebular phase. Because the oxygen that we see in the nebular phase is formed primarily in the hydrostatic burning phase of the progenitor; its mass depends sensitively on the progenitor mass and is smaller for lower-mass progenitors. Calcium, on the other hand, is synthesized by oxygen burning at a later stage and has a relatively weak dependence on progenitor mass. Thus the Ca/O ratio acts as a tracer for the progenitor mass of core-collapse SNe \citep{Fransson1989, Jerkstrand2014, Jerkstrand2015, Prentice2022, Fang2022, Ergon2022}, and a high Ca/O flux ratio implies a lower abundance of oxygen. 
The typical Type I Ca-rich transients are generally discovered in the outskirts of early-type galaxies \citep{Perets2011, Kasliwal2012, Lunnan2017}. Their old environments make core-collapse SNe unlikely to be the cause for these transients \citep{Perets2011}. However, the SNe in our sample are different from the usual Type I Ca-rich gap transients; they are found in star-forming regions and are classified as Type IIb SNe based on the photospheric spectra. 

Recently, SN 2019ehk was proposed to be a Ca-rich Type IIb SN from a low-mass stripped progenitor \citep{De2021}. This is unusual since core-collapse supernovae usually have a low [Ca~II]/[O~I] ratio. SN 2019ehk was discovered in the galaxy M100 \citep{Grzegorzek18, Jacobson2020a, Jacobson2021}. \citet{Jacobson2020a} suggested a thermonuclear explosion origin arising from a low-mass hybrid HeCO white dwarf + CO white dwarf merger. However, a core-collapse explosion from a $<$ 10 \msun\ ZAMS star could not be ruled out from archival \emph{Hubble Space Telescope} images. \citet{Nakaoka2021} suggested that SN 2019ehk originated in an ultra-stripped core-collapse SN (USSN). However, we do not expect to see hydrogen in an USSN. Thus, Ca-rich IIb events could potentially be strongly-stripped SNe (SSSNe), which is a transition class between the regular stripped-envelope SNe and ultra-stripped SNe.
        

Here, we present a holistic analysis of the lightcurve and spectral properties for a sample of nine strongly-stripped Ca-rich Type IIbs.  This work is useful to describe these potentially common types of transients, understand the fate of the low mass end of CCSNe, and determine the evolutionary pathway for Type IIb SNe. Also, this sample increases the number of Ca-rich Type IIb SNe observed in the literature from one to nine. They could provide observational constraints on SN explosions of low mass He-stars. It can also help probe the distribution of the degree of stripping, understand common envelope evolution, and constrain rates of binary mergers such as NS $-$ WD systems that could be detectable in gravitational waves.


The sample selection and data used are described in Section \ref{sec:data}. In Section \ref{sec:analysis}, we present our analysis of the late-time nebular spectra, photospheric spectra, lightcurves and the host galaxy photometry of the sample. We discuss the inferred progenitor properties, evolutionary pathways and rates in Section \ref{sec:discussion}. We provide a brief summary of the results and future goals in Section \ref{sec:conclusion}.

\section{Data}\label{sec:data}

\subsection{Sample Selection}\label{sec:sample}

In this work, \textcolor{black}{we use SNe detected by ZTF}. ZTF has a 47-square degree field-of-view camera \citep{Dekany20} which is mounted on the Palomar 48-inch (P48) Oschin Schmidt telescope. It images the entire Northern sky every $\sim$2 nights in $g$ and $r$-bands, with a median depth of around 20.5 mag. We use ZTF discoveries that are part of the Bright Transient Survey (BTS; \citealp{Fremling2020}) $-$ a magnitude-limited survey aimed at
spectroscopically classifying all SNe with peak magnitude brighter than 18.5 mag and the Census of the Local Universe survey (CLU; \citealp{De2020})$-$ a volume-limited survey aimed at classifying all SNe within 200 Mpc whose hosts belong to the CLU galaxy catalog \citep{Cook2019}.


We apply the following selection criteria on the ZTF SN sample obtained from the BTS and CLU surveys (until 31 December 2021): 

1. The transient should be classified as a Type IIb SN based on spectra obtained near peak light. We have 94 Type IIb SNe in the BTS + CLU sample.

2. We obtained at least one epoch of late-time nebular spectroscopy (at $>$ 50 rest-frame days post explosion) for events satisfying the above criteria. This was true for 29 of the 94 Type IIb SNe.

3. The [Ca~II]/[O~I] ratio in the nebular-phase spectrum should be $>$ 2 \citep{Milisavljevic2017, De2018a}. This is true for 11 SNe out of the 29 SNe for which we have nebular spectra. We excluded two SNe\footnote{SN 2019lsk, SN 2019rta} because of poor lightcurve quality. 
Table \ref{Table:sample_summary} lists the general properties of the nine SNe in the final sample.


\begin{table*} 
\begin{center} 
\caption{Summary of the sample of SNe presented in this work. Subscripts 1 and 2 refer to the peak parameters of the first peak and second peak respectively. The absolute magnitudes are corrected for Milky-Way and host-galaxy extinction (see Section \ref{sec:extinction}).} 
\begin{tabular}{ccccccccccc} 
\hline  \\  

ZTF Name & IAU Name & R.A. & Dec. & Redshift  &  $t_\mathrm{max1}$ & $M_{r1}$ &  $t_\mathrm{max2}$ & $M_{r2}$ & Av,MW & Av,host    \\ &  & (hh:mm:ss) & (dd:mm:ss) &  &  (MJD) & (mag)  & (MJD)  &  (mag) &  & \\\\ \hline  \\ 

ZTF19aatesgp  &  SN2019ehk  &  12:22:56.14  &  +15:49:34.0  &  $0.006$  &  $58606$  &  $-18.2$  &  $58616$  &  $-17.9$  &  $0.1$  &  $2.2$ \\ \hline \\ 

ZTF21aaabwfu  &  SN2021M  &  14:14:14.72  &  +35:25:23.1  &  $0.011$  &  $59216$  &  $-16.8$  &  $59231$  &  $-17.0$  &  $0.1$  &  $2.1$ \\ \hline \\ 

ZTF20abwzqzo  &  SN2020sbw  &  02:46:03.31  &  +03:19:47.6  &  $0.023$  &  $59088$  &  $-16.9$  &  $59104$  &  $-17.9$  &  $0.1$  &  $0.7$ \\ \hline \\ 

ZTF21aabxjqr  &  SN2021pb  &  09:44:46.80  &  +51:41:14.6  &  $0.033$  &  $59221$  &  $-16.9$  &  $59238$  &  $-16.8$  &  $0.0$  &  $0.0$ \\ \hline \\ 

ZTF19abxtcio  &  SN2019pof  &  01:12:37.87  &  +33:02:05.7  &  $0.015$  &  $58733$  &  $-15.6$  &  $58749$  &  $-16.7$  &  $0.2$  &  $0.5$ \\ \hline \\ 

ZTF18abwkrbl  &  SN2018gjx  &  02:16:15.58  &  +28:35:28.6  &  $0.012$  &  $58379$  &  $-17.8$  &  $58390$  &  $-16.9$  &  $0.2$  &  $0.0$ \\ \hline \\ 

ZTF19abacxod  &  SN2019hvg  &  14:06:01.58  &  +12:46:50.3  &  $0.018$  &  $58654$  &  $-16.6$  &  $58672$  &  $-17.6$  &  $0.1$  &  $0.8$ \\ \hline \\ 

ZTF18acqxyiq  &  SN2018jak  &  09:59:18.19  &  +34:53:43.7  &  $0.038$  &  $58447$  &  $-18.1$  &  $58467$  &  $-17.5$  &  $0.0$  &  $0.6$ \\ \hline \\ 

ZTF21abjyiiw  &  SN2021sjt  &  20:37:19.19  &  +66:06:23.1  &  $0.005$  &  $59403$  &  $-16.9$  &  $59417$  &  $-16.7$  &  $1.0$  &  $1.9$ \\ \hline \\ 

\end{tabular}  \label{Table:sample_summary} 
\end{center} 
\end{table*}

\subsection{Optical photometry}
\label{section:photdata}

\begin{figure*}
    \centering
    \includegraphics[width=17.5cm]{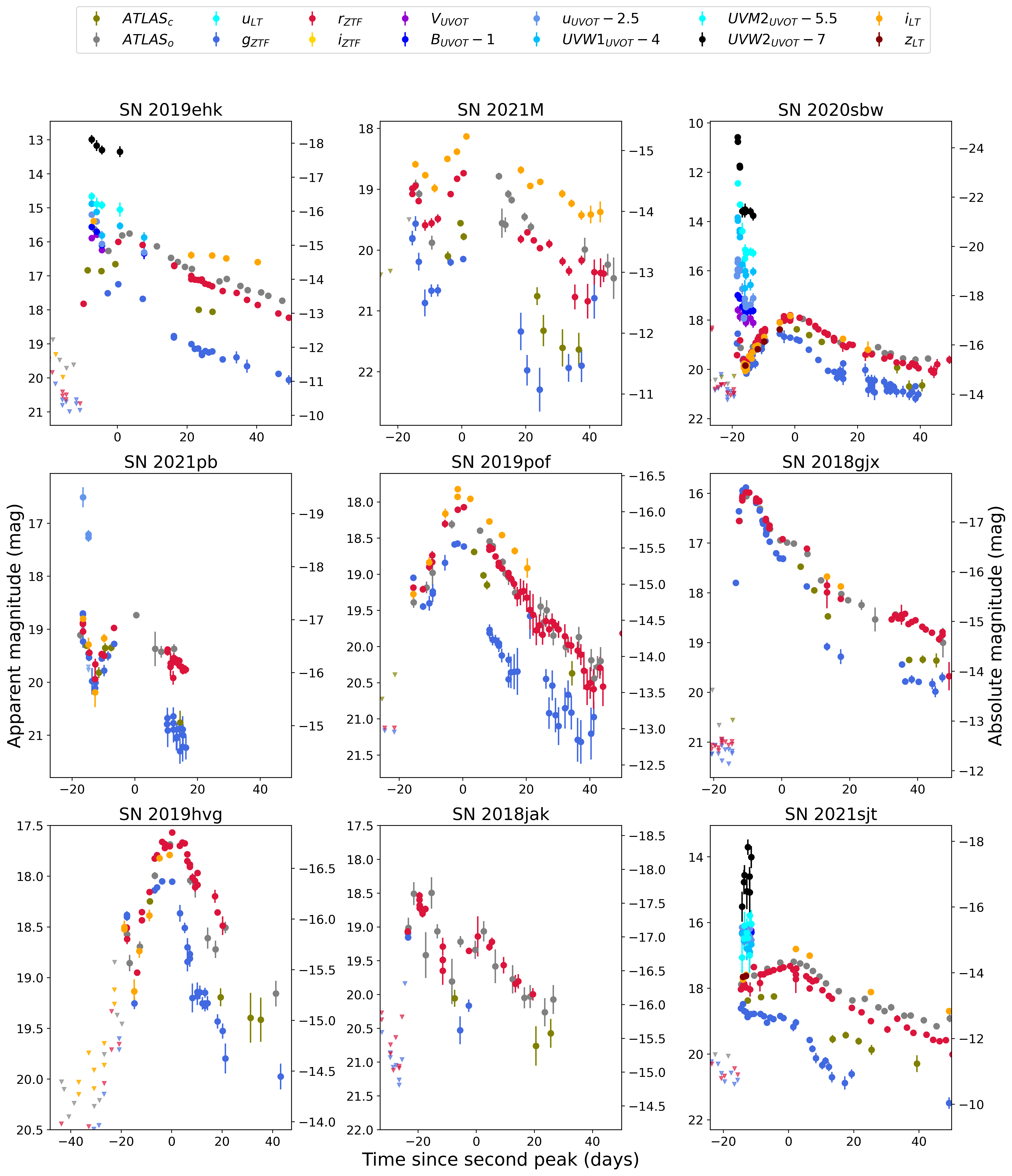}
    \caption{Lightcurves for our sample of nine Ca-rich Type IIb SNe. This has forced-photometry from ZTF, ATLAS and follow-up observations from various instruments. See Section \ref{section:photdata} for further description on the photometry. The left y-axis shows the apparent magnitude (mag) and the right y-axis shows the absolute magnitude (mag), without host extinction correction. The x-axis shows the number of rest-frame days since the second peak.}
    \label{fig:all_lc}
\end{figure*}

We perform forced PSF photometry on the ZTF difference images via the ZTF forced-photometry service
\citep{Masci2019} in $g$, $r$ and $i$ bands. For this work, we consider
anything less than a 3$\sigma$ detection an upper limit. The ZTF lightcurves were supplemented with photometry from the Palomar 60-inch telescope \citep[P60; ][]{Cenko2006}, the Optical wide field
camera on the Liverpool Telescope \citep[LT; ][]{Steele2004}, Sinistro and
SBIG imagers on the 1-meter class and the Spectral imager
on the 2-meter class telescopes operated by Las Cumbres Observatory
\citep[LCO; ][]{Brown2013} in $g$, $r$ and $i$ bands. We also use photometry in $u$, $i$, $z$ bands obtained with the LT for a few sources. These data were processed with the automatic image subtraction pipeline FPipe
\citep{Fremling2016} using reference images from the Sloan Digital Sky Survey (SDSS; \citealt{Ahn2012a})
and PanSTARRS \citep[PS1; ][]{Chambers2016}. We have early-time UV data from the Ultra-violet Optical Telescope (UVOT; \citealt{Roming2005}), which is on board the  \emph{Neil Gehrels Swift Observatory}  \citep{Gehrels2004}. The UVOT data is reduced using \HEAsoft\footnote{\href{https://heasarc.gsfc.nasa.gov/docs/software/heasoft}{https://heasarc.gsfc.nasa.gov/docs/software/heasoft}}. We also use forced-photometry data points from the Asteroid Terrestrial-impact Last Alert System \citep[ATLAS; ][]{Tonry2018, Smith2020}. Figure \ref{fig:all_lc} shows a multi-paneled figure of the lightcurves for the SNe in our sample.

\subsection{Optical spectroscopy}
\label{section:spectra}

\begin{figure*}
    \centering
    \includegraphics[width=14.5cm]{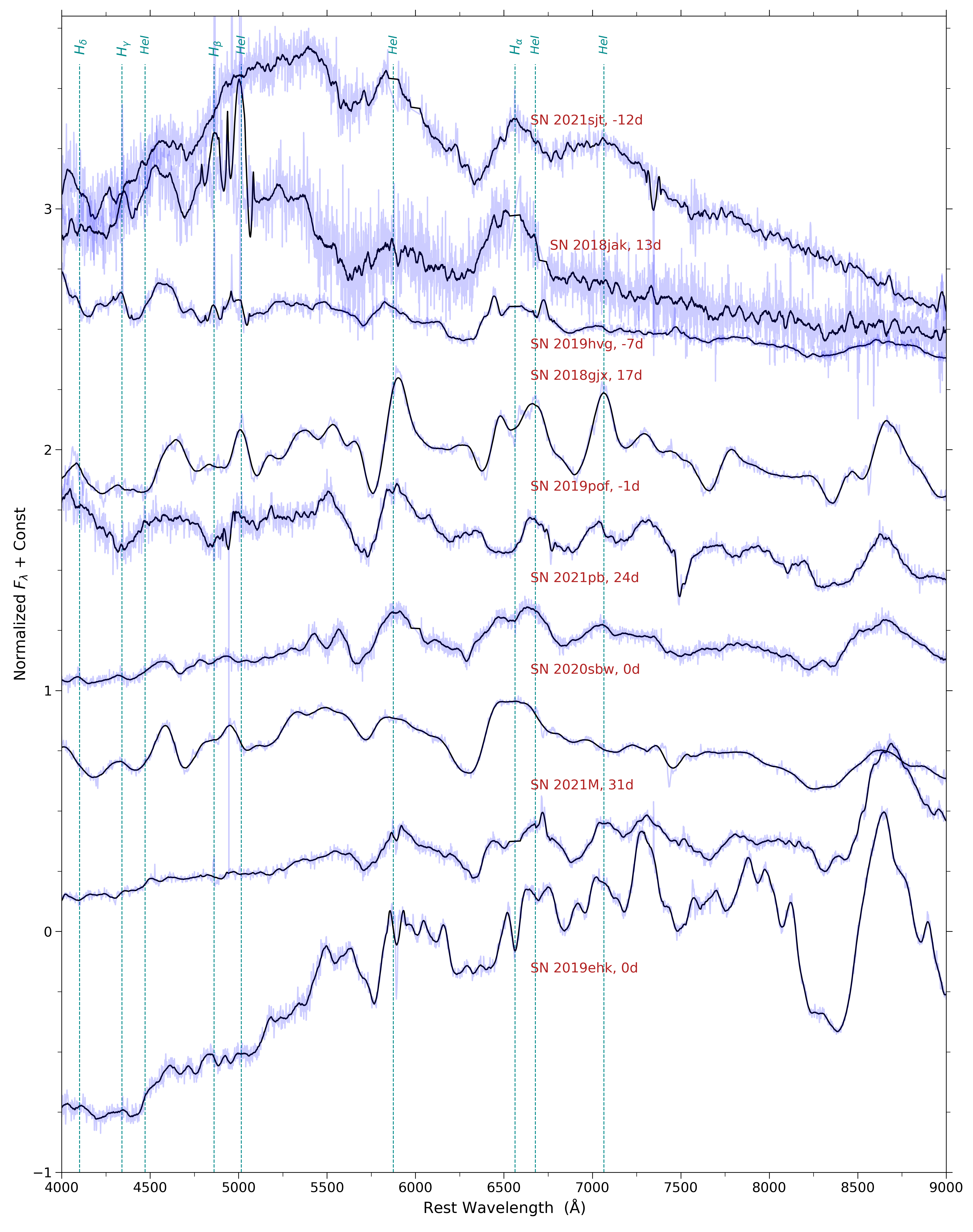}
    \caption{Collage of few spectra for the SNe in our sample showing H and He emission lines. The phase of the spectra is measured as the number of rest-frame days since the second peak.}
    \label{fig:all_photspectra}
\end{figure*}

\begin{figure*}
    \centering
    \includegraphics[width=7.5cm]{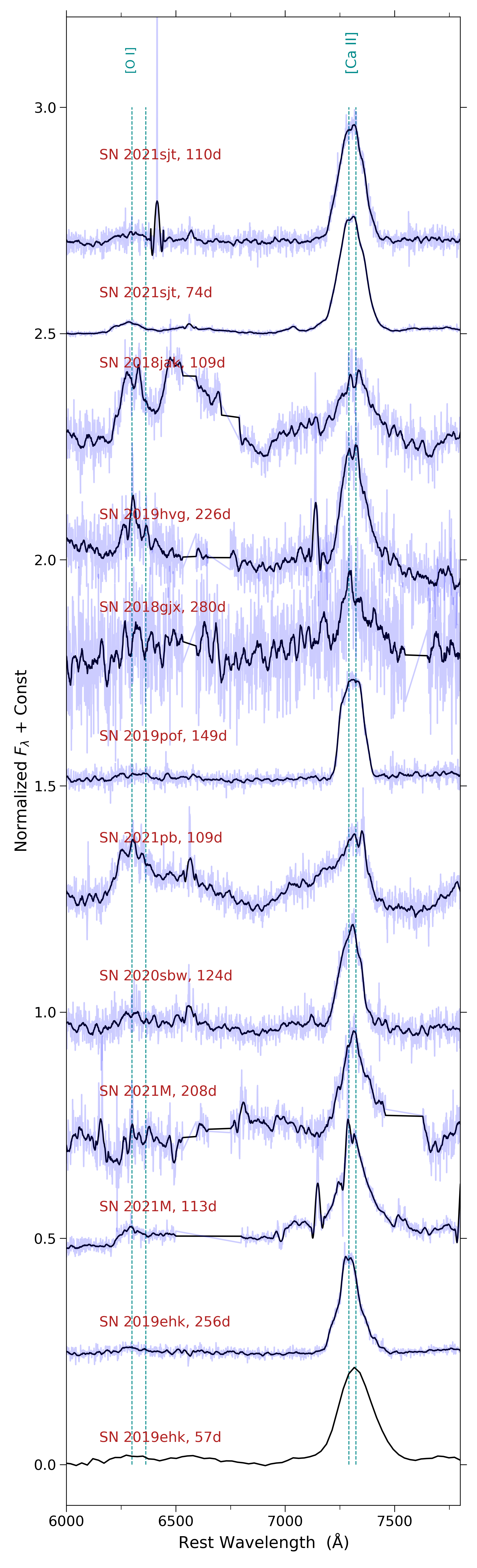}
    \caption{Nebular spectra showing the [Ca~II] $\lambda \lambda$7291, 7324 and  [O~I]  $\lambda \lambda$6300, 6364 lines. The phase of the spectra is measured as the number of rest-frame days since the second peak.} The spectra with the full wavelength range are shown in Appendix \ref{appendix:nebfull}.
    \label{fig:all_nebspectra}
\end{figure*}

We obtained spectroscopy at multiple
epochs for the sources in our sample, with phases varying from one day to over 200 days past the explosion day. One spectrum is obtained near peak luminosity for initial spectroscopic classification. Further spectral follow-up is part of the classification effort of the BTS and CLU surveys. We primarily use the Double Beam Spectrograph \citep[DBSP; ][]{Oke1982} on the Palomar 200-inch telescope and the Spectral Energy Distribution Machine \citep[SEDM; ][]{Blagorodnova2018, Rigault2019} on the Palomar 60 inch telescope for classification. The DBSP spectra reduction pipelines used are described in \citet{Bellm2016} and \citet{Roberson2022}. The SEDM data reduction pipeline is described in \citet{Rigault2019}. We also use spectra from the Spectrograph for the Rapid Acquisition of Transients \citep[SPRAT;][]{Piascik2014} on the Liverpool Telescope, the Alhambra Faint Object Spectrograph and Camera (ALFOSC) on the Nordic Optical Telescope \citep[NOT; ][]{Djupvik2010} and the Device Optimized for the LOw RESolution (DOLORES) on the Telescopio Nazionale Galileo (TNG). The SPRAT pipeline is based on the FrodoSpec pipeline \citep{Barnsley2012}. The NOT and TNG data were reduced using the data-reduction pipelines PyNOT\footnote{\href{https://github.com/jkrogager/PyNOT}{https://github.com/jkrogager/PyNOT}} and PypeIt \citep{Prochaska2020a}. Late-time nebular-phase spectra were obtained using the Low Resolution Imaging Spectrometer (LRIS) on the Keck I telescope starting $\approx$ 50 days post explosion. For a couple of sources, we obtained multiple epochs of nebular-phase spectra using LRIS. The LRIS spectra were reduced using the automated \lpipe{} \citep{Perley2019} pipeline. We used a publicly available LRIS spectrum for SN 2021sjt from 8 July 2021 \citep{Siebert2021}. Table \ref{table:vel} lists the log of the observed spectra. Figure \ref{fig:all_photspectra} shows one photospheric spectrum obtained for each source and Figure \ref{fig:all_nebspectra} shows the obtained nebular spectra.

\subsection{Host photometry} 

We obtain science-ready co-added images from the \textit{Galaxy Evolution Explorer} (\galex) general release 6/7 \citep{Martin05}, Sloan Digital Sky Survey data release 9 (SDSS DR 9; \citealt{Ahn2012a}), the Two Micron All Sky Survey \citep[2MASS;][]{Skrutskie2006a}, the Panoramic Survey Telescope and Rapid Response System (Pan-STARRS, PS1) Data Release 1 \citep{Chambers2016}, and preprocessed \wise\ images \citep{Wright2010a} from the unWISE archive \citep{Lang2014a}\footnote{\href{http://unwise.me}{http://unwise.me}}. The unWISE images are based on the public \wise\ data and from the ongoing NEOWISE-Reactivation mission R3 \citep{Mainzer2014a, Meisner2017a}. We measure the brightness of the host galaxies using LAMBDAR\footnote{\href{https://github.com/AngusWright/LAMBDAR}{https://github.com/AngusWright/LAMBDAR}} \citep[Lambda Adaptive Multi-Band Deblending Algorithm in R;][]{Wright2016a} and the methods described in \citet{Schulze2020a}. We use the offsets reported by \citet{Blanton2007a} and \citet{Cutri2013a} to convert the 2MASS and unWISE photometry from the Vega system to the AB system.

\section{Methods and Analysis} \label{sec:analysis}

\subsection{Extinction correction}
\label{sec:extinction}


We need an estimate of the extinction of the SN light due to dust along the line-of-sight to the observer in order to accurately estimate its explosion parameters. Extinction is typically split into two components. The first Galactic component accounts for extinction due to dust within the Milky Way and the second component accounts for extinction external to the Milky Way, mostly attributed to the host galaxy of the SNe. We correct for Galactic extinction by using the reddening maps of \citet{Schlafly11}. We use the \citet{Cardelli1989} extinction law and $R_V = 3.1$ for reddening corrections. Accounting for extinction external to our own galaxy is more challenging. In this paper, we use two different approaches to correct for the host-extinction. 

First, we use the intrinsic color template from the Carnegie Supernova Project (CSP-I) \citep{Stritzingetr2018}. This assumes that Type IIb SNe show similar intrinsic colors after peak.  We measure the average difference of our observed colors and the intrinsic template between 0 and 20 days after $r$-band peak as the excess color due to host extinction. We use $R_V = 3.1$ and \citet{Cardelli1989} extinction law to correct for the reddening. Shock-cooling breakout modeling (described in Section \ref{section:lcfit}) provides an opportunity to put an upper limit to the host extinction value. When we use host extinction using the CSP template for SN 2021sjt, we get an unphysically large progenitor envelope radius ($R_{env}$) of around $1.7 \times 10^6\ \Rsun$. Thus, we need to be cautious while using the intrinsic template given by \citet{Stritzingetr2018} based on a sample size of three Type IIb SN.

Second, we use the narrow host-galaxy absorption lines of Na I D. This assumes that there is a relation between host-galaxy sodium absorption lines and the amount of extinction \citep{Poznanski2012, Stritzingetr2018}. 
To compute the Na I D Equivalent Width (EW), we use spectra obtained with DBSP, ALFOSC, and LRIS, where the signal-to-noise of the spectrum allows the measurement of the EW. To compute $A_V$, we use $A_V^{host} [\rm mag] =0.78 (\pm 0.15) \times EW_{\rm Na\ I\ D} [\mathrm{\AA}]$ reported in \citet{Stritzingetr2018}. We note the Na I D EW measurement is not always a reliable tracer of the host extinction. This is particularly true for low-resolution spectra \citep{Poznanski2011} and in the presence of circumstellar material \citep{Phillips2013}.

The extinction values measured as described are shown in Table \ref{Table:hostext}. For the lightcurve analysis done in this paper, we use the average of the extinction values we get from these methods. We show the color evolution of our objects \textcolor{black}{compared to the Type IIb SNe sample of \citet{Taddia2018} in Figure \ref{fig:color}.}

\begin{figure}[ht]
    \centering
    \includegraphics[width=1.1\columnwidth]{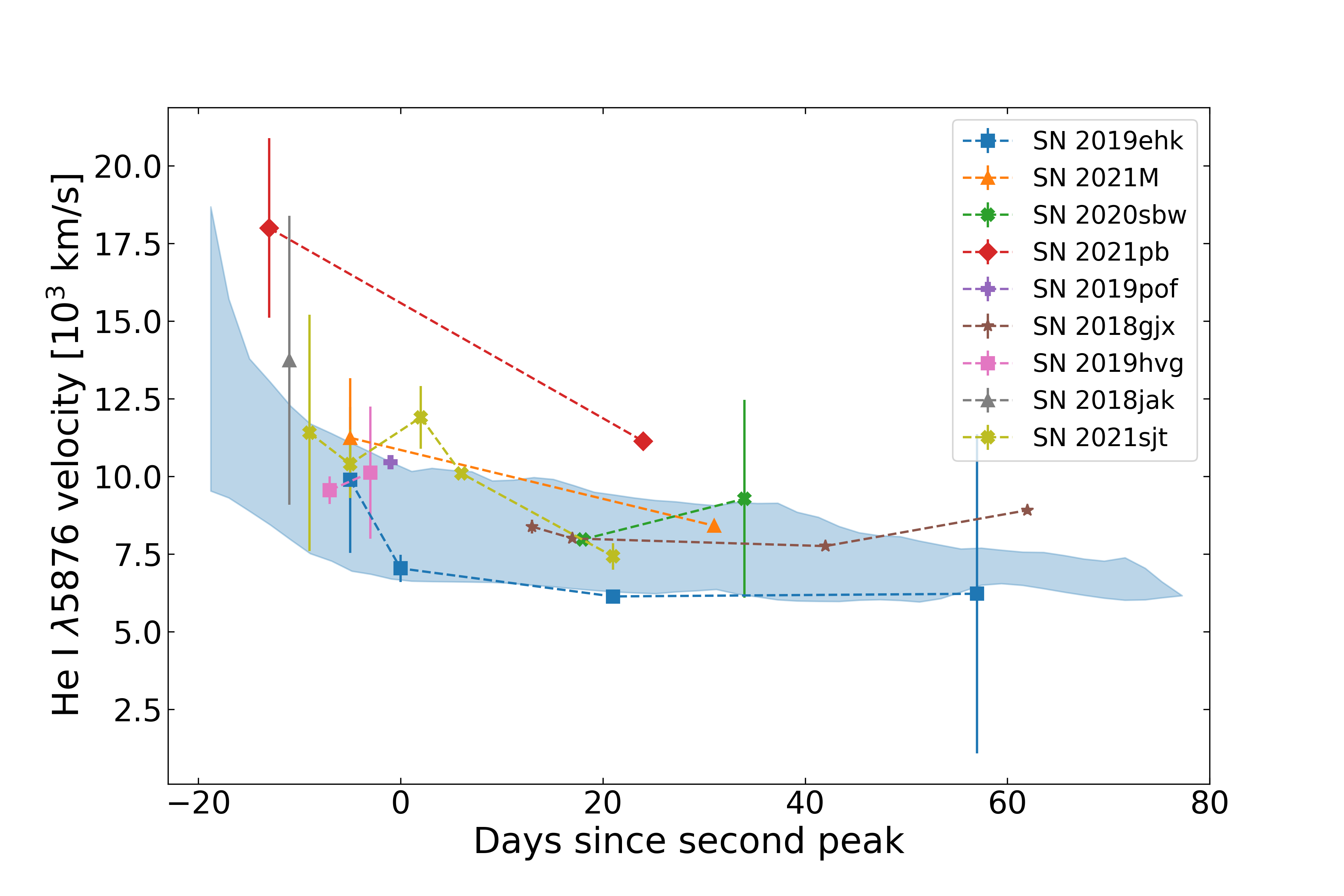}
    \caption{The expansion velocity of the He I line for each SN of our sample represented by the filled dots. The blue shaded region shows the  $1\sigma$ region of the ejecta velocity calculated for a sample of usual Type IIb SNe in \citet{Fremling2018}}
    \label{fig:ejectavel}
\end{figure}

\subsection{Photospheric spectra}
\label{sec:photspectra}

As described in Section \ref{section:spectra}, we obtain spectra close to the peak epoch for all the sources. Classifications were done using the \texttt{SuperNova Identification} \citep[\texttt{SNID};][]{Blondin2007} code. We used the \texttt{superfit} \citep{Howell2005} code to classify spectra contaminated by the host galaxy. The final classification as Type IIb was made after manual inspection of the emission and absorption lines and the best-fit templates matched from \texttt{SNID} or \texttt{superfit}. 

  
We determine the expansion velocities of the H~I $\lambda 6563$, He~I $\lambda 5876$, O~I $\lambda 7774$ lines where absorption from these lines could be identified and measurements were possible. To do this, we fit the minima of the absorption part of the P-Cygni profiles of the spectral lines with a polynomial fit (with the polynomial degree manually tuned for each spectrum, but typically 3). The minima of the absorption features are then used to estimate the expansion velocity. For certain cases, where the spectra are contaminated with galaxy lines or the resolutions is not high, we find the minima of the absorption portion by manual inspection. We list the velocities obtained in Table \ref{table:photspectra}. To estimate uncertainties on our measurements, we use a Monte-Carlo approach. We obtain a noise spectrum by subtracting a heavily smoothed version of the spectrum from the original spectrum. To get an estimate of the noise, we then calculate the standard deviation of this noise spectrum. Then, we create simulated noisy spectra by adding noise using a standard gaussian distribution with the standard deviation calculated above. To the heavily smoothed spectra, we add these simulated noise spectra and re-calculate the velocity from the absorption minima for each simulated spectrum.   The standard deviation of these measurements is taken as the 1$\sigma$ uncertainty of our velocity measurements.  
\citet{Fremling2018} analyzed the spectra of a sample of 55 Type IIb SNe discovered by the Palomar Transient Factory (PTF) and intermediate PTF (iPTF) surveys. We compare the velocities we have with the mean velocities obtained in \citet{Fremling2018} for the He~I $\lambda 5876$ line. From Figure \ref{fig:ejectavel}, we find that the expansion velocities obtained using the He~I $\lambda 5876$  line are consistent with those of typical Type IIb SNe obtained by \citet{Fremling2018}. 

\subsection{Nebular spectra}
\label{section:nebspectra}

We obtained late-time photometry using the LCO, Keck, P60 and P48 telescopes as recorded in Table \ref{table_neb}. Since the [Ca II] line lies entirely in the $i$-band, the interpolated late-time photometry is used to flux calibrate our nebular spectra. When late-time photometry is not available, we extrapolate the lightcurve by assuming a late-time ($>$ 30 days) $i$-band decline rate of $0.019\ \pm\ 0.004$ mag/day, based on the average late-time decay of the SESNe tabulated in \citet{Wheeler2015}.

We measure the line fluxes by numerical integration of the respective wavelength regions. We add noise (scaled to nearby regions of the continuum) to the line profile and use Monte Carlo sampling of the estimated fluxes to quantify the uncertainties. The measured [Ca~II] $\lambda \lambda$7291, 7324 [O~I] and $\lambda \lambda$6300, 6364 fluxes and their ratio are shown in Table \ref{table_neb}.



\subsection{Modeling lightcurves}\label{sec:bbfit}

\subsubsection{Blackbody fit}

We estimate the bolometric lightcurve at epochs where at least detections in two filters are available by fitting a blackbody function. For each epoch, we do the blackbody fit using a Markov Chain Monte-Carlo (MCMC) method using the Python \emcee\ package \citep{Foreman-Mackey13} to estimate the temperature, radius and bolometric luminosity. We use the  $16^{th}$ and $84^{th}$ percentiles of the posterior probability distribution as the model uncertainties. The blackbody fits for each source are available in Appendix \ref{appendix:bbfit}.

\begin{table*}
\begin{center}
\caption{Summary of the host extinction parameters for our sample. } 

\begin{tabular}{cccccccc} 
\hline \\
Source &  Telescope  & Date &  Na I D EW & Avg. Color  ($g-r$) & $A_\mathrm{V,\ Na I D}$  & $A_\mathrm{V,\ color}$ & $A_\mathrm{V}$ used \\ & + Instrument &  &   (\AA) & (mag)  & (mag) & (mag) & (mag)  \\\\
\hline \\

SN 2019ehk & P200+DBSP    & 13 May 2019 &  $2.99 \pm\ 0.27$ & $0.68 \pm\ 0.03$  & $2.3$(0.2) & 2.3(0.2) &  2.2\\\\
& Keck1+LRIS    & 3 June 2019 & $2.55 \pm\ 0.43$ & & 2.0(0.3)  & \\

\hline \\

SN 2021M & Keck1+LRIS    & 12 Jan 2021  & $2.03 \pm\ 0.22$ & $0.95 \pm\ 0.4$ & $1.6$(0.2)  & 3.2(0.3) &  2.1\\\\

 & Keck1+LRIS    & 17 Feb 2021 & $2.37 \pm\ 0.21$ && 1.8(0.2) &\\\\

 & Keck1+LRIS    & 10 May 2021 & $2.12 \pm\ 0.24$ && 1.7(0.2) &\\\\

 & Keck1+LRIS    & 13 Aug 2021   &  $2.71 \pm\ 0.41$ && 2.1(0.3) &\\
 \hline \\
 
SN 2020sbw & NOT+ALFOSC  &  14 Sep 2020  & $< 0.1$  & $0.51 \pm\ 0.10$  & $< 0.1$ & 1.4(0.1) & 0.7 \\

\hline \\

SN 2021pb & -  &  &   &      & - & - &  0 \\

\hline \\

SN 2019pof & -  &  &   &  $0.38 \pm\ 0.11$    & -  & 0.5(0.1) &  0.5 \\

\hline \\

SN 2018gjx & -  &  &   &      & - & - &  0\\

\hline \\

SN 2019hvg & -  &  &   &   $0.59 \pm\ 0.21$   & - & 0.8(0.1) & 0.8 \\

\hline \\

SN 2018jak & Keck1+LRIS    & 4 Dec 2018  & $0.8 \pm\ 0.35$ & &  $0.6$(0.3)  & - & 0.6\\

\hline \\

SN 2021sjt & Keck1+LRIS & 8 July 2021  & $2.31 \pm\ 0.15$ & $1.1 \pm\ 0.2$ & $1.8$(0.1)  & $3.3$(0.3) &  1.9\\\\

& P200+DBSP & 10 July 2021 & $2.39 \pm\ 0.19$ & & 1.9(0.1) & & \\\\

& P200+DBSP & 07 Aug 2021  & $1.57 \pm\ 0.24$ & & 1.2(0.2) && \\\\

& Keck1+LRIS  & 12 Aug 2021 & $1.50 \pm\ 0.31$ & & 1.2(0.2) && \\


\hline \\

\end{tabular} \label{Table:hostext}
\end{center}
\end{table*}

\begin{table*} 
\begin{center}
\caption{Log of the spectral observations and expansion velocity for few spectral lines in km/s. The phases are measured as the number of rest-frame days since the second peak.} \label{table:photspectra}
\begin{tabular}{ccccccccc} 

\hline \\
Object &  Instrument & Range & Date & Phase   & H I 6563 & He I 5876 & O I 7774 \\\\
&& [$\rm \AA$] & & [days] & [km/s] & [km/s] & [km/s]  \\\\
\hline \\

SN 2021M & P60+SEDM & 3500-9220 & 10 Jan 2021 & -7d & $18100 \pm\ 4200$ & - & - \\\\ 
& Keck 1+LRIS&  3200-10 000 & 12 Jan 2021 & -5d &  $14500 \pm\ 300$ &  $11000 \pm\ 1900$ & - \\\\
& Keck 1+LRIS &  3200-10 000 & 17 Feb 2021 & +31d & $12400 \pm\ 100$ & $8400 \pm\ 200$  & $3700 \pm\ 300$ \\\\
& Keck 1+LRIS&  3200-10 000 &  10 May 2021 & +113 & -       & $9100 \pm\ 2700$ & $4700 \pm\ 1600$ \\\\
& Keck 1+LRIS&  3200-10 000 &  13 Aug 2021 & +208 & -  & -    & $2900 \pm\ 900$ \\
\hline \\

SN 2020sbw & P200+DBSP & 3200-10 100 & 29 Aug 2020& -14d & - & - & - \\\\
& P60+SEDM & 3500-9220 & 12 Sep 2020& 0d & $13400 \pm\ 1000$ & & & \\\\
& NOT+ALFOSC & 3974-9153 & 14 Sep 2020& +2d & $13200 \pm\ 100$ & $7900 \pm\ 100$ & - & -  \\\\
& P60+SEDM & 3500-9220 & 30 Sep 2020& +18d & $11400 \pm\ 4100$ & $9300 \pm\ 3200$ & -	 \\\\
& P60+SEDM & 3500-9220 & 7 Oct 2020& +25d & - & - & -	 \\\\
& NOT+ALFOSC & 3974-9153 & 16 Oct 2020& +34d & $13100 \pm\ 200$ & $8900 \pm\ 200$ & - \\\\
& Keck 1+LRIS &  3200-10 000 & 14 Jan 2021 & +124d & - & - & - & \\
\hline \\

SN 2019pof & P60+SEDM & 3500-9220 & 18 Sep 2019  & \textcolor{black}{-5d} & - & - &- \\\\
& P200+DBSP & 3200-10 100 & 22 Sep 2019  & \textcolor{black}{-1d} & $15200 \pm\ 600$ & $10400 \pm\ 200$ &- \\\\
& Keck 1+LRIS &  3200-10 000 & 19 Feb 2020 & \textcolor{black}{+149d} & - & - & - & \\

\hline \\

SN 2019hvg & P60+SEDM & 3500-9220 & \textcolor{black}{20 June 2019}  & 
\textcolor{black}{-18d} & $18000 \pm\ 4500$ & - & -  \\\\
& P60+SEDM & 3500-9220 & 24 June 2019  & \textcolor{black}{-14d} & $16200 \pm\ 3900$ & - & -  \\\\
&  P200+DBSP & 3200-10 100 & 1 July 2019   & \textcolor{black}{-7d} & $13200 \pm\ 400$ &	$9500 \pm\ 400$ &	- \\\\
& P60+SEDM & 3500-9220 & 5 July 2019  & \textcolor{black}{-3d} &  $12400 \pm\ 900$ &	$10100 \pm\ 2100$	& -  \\\\
& Keck 1+LRIS &  3200-10 000 & 19 Feb 2020 & \textcolor{black}{+226d} & - & - & - & \\

\hline \\

SN 2019ehk & P60+SEDM & 3500-9220 & 8 May 2019 & \textcolor{black}{-5d} &  $16500 \pm\ 3200$ &	$9900 \pm\ 2300$	& - 	 \\\\
& P200+DBSP & 3200-10 100 & 13 May 2019  & \textcolor{black}{0d} & - & $7039 \pm\ 400$ &	 -  \\\\
&  Keck 1+LRIS &  3200-10 000 & 3 June 2019 & \textcolor{black}{+21d} & - &	$6100 \pm\ 100$ &	 -  \\\\

\hline \\

\end{tabular} \label{table:vel}
\end{center}
\end{table*}    

\begin{table*}
\ContinuedFloat
\begin{center}
\caption{Continued ...} 

\begin{tabular}{cccccccc} 

\hline \\
Object &  Instrument & Range & Date & Phase   & H I 6563 & \textcolor{black}{He I 5876} & O I 7774 \\\\
&& [$\rm \AA$] & & [days] & [km/s] & [km/s] & [km/s]  \\\\
\hline \\
&  P60+SEDM &  3200-10 000 & 9 July 2019 & \textcolor{black}{+57d} & - &	$6200 \pm\ 5100$ &	 -  \\\\
&  Keck 1+LRIS &  3200-10 000 & 24 Jan 2020 & \textcolor{black}{+256d} & - &	- &	 -  \\\\

\hline \\

SN 2021pb &  P60+SEDM & 3200-10 000 & 7 Jan 2021  & \textcolor{black}{-17d} &  - & - & -\\\\
& Keck 1+LRIS &  3200-10 000 & 11 Jan 2021 & \textcolor{black}{-13d} & $18400 \pm\ 2900$ & $17900 \pm\ 2900$ & - \\\\
 & Keck 1+LRIS &  3200-10 000 & 17 Feb 2021 & \textcolor{black}{+24d} & $14300 \pm\ 1800$ & $11100 \pm\ 200$ & $9100 \pm\ 800$  \\\\
 & Keck 1+LRIS &  3200-10 000 & 13 May 2021 & \textcolor{black}{+109d} & - & $7900 \pm\ 2900$ & $6100 \pm\ 2000$ \\
\hline \\

SN 2018gjx  &  P60+SEDM & 3200-10 000 & \textcolor{black}{18  Sep 2018}  & \textcolor{black}{-11d} &  - & - & -\\\\
& NTT+EFOSC2  & 3200-10 000 & 18 Sep 2018  & \textcolor{black}{-11d} & - & - & - \\\\
& P60+SEDM    & 3200-10 000 &12 Oct 2018  &  \textcolor{black}{+13d} & $11100 \pm\ 200$ & $8400 \pm\ 200$ & $4000 \pm\ 600$ \\\\
& NOT+ALFOSC  & 3974-9153 &  16 Oct 2018  & \textcolor{black}{+17d} & $8300 \pm\ 100$ & $8000 \pm\ 100$ & $3800 \pm\ 100$ \\\\
& Keck1+LRIS  & 3200-10 000 &10 Nov 2018  & \textcolor{black}{+42d} & $9400 \pm\ 100$ & $7800 \pm 100$   & $4300 \pm\ 100$ \\\\
& TNG+DOLORES & 3200-10 000 & 30 Nov 2018  & \textcolor{black}{+62d} & $10500 \pm\ 600$ & $8900 \pm\ 100$  & $7000 \pm\ 400$ \\\\
& Keck1+LRIS  & 3200-10 000 &5 Jan 2019 &  \textcolor{black}{+98d} & $15100 \pm\ 3400$ & $9300 \pm\ 400$ & $5500 \pm\ 300$ \\\\
& Keck1+LRIS  & 3200-10 000 &6 July 2019 & \textcolor{black}{+280d} & - & - & - \\
\hline \\

SN 2018jak & Keck1+LRIS  &   3200-10 000 & 4 Dec 2018  & \textcolor{black}{-11d} & $15800 \pm\ 1700$ & $13700 \pm\ 4600$ & - \\\\
& P60+SEDM &  3200-10 000 & 28 Dec 2018  & \textcolor{black}{+13d} & $11000 \pm\ 8400$ & - & - \\\\
& Keck1+LRIS  &  3200-10 000 & 3 April 2019 & \textcolor{black}{+109d} & $7500 \pm\ 1200$ & $6600 \pm\ 1500$ & - \\
\hline \\


\end{tabular} 
\end{center}
\end{table*}

\begin{table*}
\ContinuedFloat

\begin{center}
\caption{Continued ...} 

\begin{tabular}{cccccccc} 

\hline \\
Object &  Instrument & Range & Date & Phase   & H I 6563 & He I 5876 & O I 7774 \\\\
&& [$\rm \AA$] & & [days] & [km/s] & [km/s] & [km/s]  \\\\
\hline \\

SN 2021sjt & Keck1+LRIS  & 3200-10 000 & 8 July 2021 & \textcolor{black}{-14d} & $10600 \pm\ 1000$ & - &  -\\\\
& P60 + SEDM & 3200-10 000 & 8 July 2021 & \textcolor{black}{-14d} & - & - & - \\\\
& NOT + ALFOSC & 3974-9153 & 8 July 2021 & \textcolor{black}{-14d} & $11600 \pm\ 3300$ & - & - \\\\
& P200+DBSP  & 3200-10 100 & 10 July 2021 & \textcolor{black}{-12d} & $9900 \pm\ 7600$ & - & $6500 \pm\ 3100$ \\\\
& P60 + SEDM & 3200-10 000 &10 July 2021 & \textcolor{black}{-12d} & - & - & - \\\\
& P60 + SEDM & 3200-10 000 &13 July 2021 & \textcolor{black}{-9d}  & - & $11400 \pm\ 3800$ & $6800 \pm\ 3200$ \\\\

& LT + SPRAT & 3200-10 000 & 17 July 2021 & \textcolor{black}{-5d} & $5900 \pm\ 1100$ & $10400 \pm\ 1100$ & $6300 \pm\ 100$ \\\\

& P60 + SEDM & 3200-10 000 &24 July 2021 & \textcolor{black}{+2d} & - & $11900 \pm\ 7300$ & - \\\\

& NOT + ALFOSC & 3974-9153 &28 July 2021 & \textcolor{black}{+6d} & - & $10100 \pm\ 100$ & $5900 \pm\ 200$ \\\\

& P60 + SEDM & 3200-10 000 &28 July 2021 & \textcolor{black}{+6d} & -   & $12500 \pm\ 3800$ & $7400 \pm\ 2300$ \\\\
& P200+DBSP  & 3200-10 100 &7 Aug 2021 & \textcolor{black}{+16d} & $6900 \pm\ 700$ &  - & $4300 \pm\ 900$\\\\
& Keck1+LRIS  & 3200-10 000 &12 Aug 2021 & \textcolor{black}{+21d} & $7400 \pm\ 800$ & $7400 \pm\ 200$ & - \\\\ 
& P60 + SEDM & 3200-10 000 &18 Aug 2021 & \textcolor{black}{+27d} & - & - & -\\\\
& P60 + SEDM & 3200-10 000 &23 Sep 2021 & \textcolor{black}{+63d} & - & - & - \\\\
& Keck1+LRIS  & 3200-10 000 &4 Oct 2021 & \textcolor{black}{+74d} & - & - & - \\\\ 
& Keck1+LRIS  & 3200-10 000 &9 Nov 2021 & \textcolor{black}{+110d} & - & - & - \\\\ 
\hline \\

\end{tabular} 
\end{center}
\end{table*} 

\begin{table*} 
\begin{center} 
\caption{Summary of the nebular properties. \textcolor{black}{The phases are measured as the number of rest-frame days since the second peak. We note that the oxygen mass measured here is the lower limit on the oxygen mass in the ejecta. The lower and upper limits of the O mass provided in the `O mass' column correspond to 4000 K and 3400 K, respectively.}} 
\begin{tabular}{ccccccc} 
\hline  \\  

Source &  Spectra Tel.+Inst./Phase   &  Phot. Tel.+Inst./Phase & Phot. mag & [Ca II]/[O I] flux ratio & [O I] lum. & O mass   \\    & (days)   & (days)  & (mag)  & & ($10^{38} \mathrm{erg s^{-1}}$) & \textcolor{black}{($10^{-2}$ \Msun)}\\ \hline  \\


SN 2019ehk  &  Keck1+LRIS/\textcolor{black}{256}  &  Keck1+LRIS/\textcolor{black}{280}  &  $22.10 \pm 0.15$  &  $28.58 \pm 3.19$  &  $1.72 \pm 0.17$  &  \textcolor{black}{$3.86 - 12.91$}  \\ \hline \\ 

SN 2021M  &  Keck1+LRIS/\textcolor{black}{113}  &  LCOGT 1m/\textcolor{black}{176}  &  $21.75 \pm 2.19$  &  $3.26 \pm 1.84$  &  $0.15 \pm 0.01$  &  \textcolor{black}{$0.35 - 1.12$}  \\ \hline \\ 

SN 2021M  &  Keck1+LRIS/\textcolor{black}{208}  &  LCOGT 1m/\textcolor{black}{176}  &  $21.75 \pm 2.19$  &  $4.32 \pm 0.22$  &  $0.93 \pm 0.03$  &  \textcolor{black}{$2.25 - 6.58$}  \\ \hline \\ 

SN 2020sbw  &  Keck1+LRIS/\textcolor{black}{124}  &  LCOGT 1m/\textcolor{black}{115}  &  $>20.63$  &  $10.37 \pm 4.65$  &  $<11.69$  &  \textcolor{black}{$<17.29$}  \\ \hline \\ 

SN 2021pb  &  Keck1+LRIS/\textcolor{black}{109}  &  LCOGT 1m/\textcolor{black}{27}  &  $>20.21$  &  $1.73 \pm 0.04$  &  $<6.32$  &  \textcolor{black}{$<15.58$}  \\ \hline \\ 

SN 2019pof  &  Keck1+LRIS/\textcolor{black}{149}  &  P48/\textcolor{black}{20}  &  $19.02 \pm 0.30$  &  $32.37 \pm 1.47$  &  $0.01 \pm 0.00$  &  \textcolor{black}{$0.04 - 0.10$}  \\ \hline \\ 

SN 2018gjx  &  Keck1+LRIS/\textcolor{black}{102}  &  P60/\textcolor{black}{98}  &  $19.61 \pm 0.21$  &  $2.24 \pm 0.51$  &  $0.08 \pm 0.02$  &  \textcolor{black}{$0.16 - 0.64$}  \\ \hline \\ 

SN 2019hvg  &  Keck1+LRIS/\textcolor{black}{226}  &  P48/\textcolor{black}{43}  &  $19.35 \pm 0.19$  &  $3.69 \pm 0.48$  &  $0.11 \pm 0.01$  &  \textcolor{black}{$0.25 - 0.79$}  \\ \hline \\ 

SN 2018jak  &  Keck1+LRIS/\textcolor{black}{109}  &  P48/\textcolor{black}{109}  &  $>20.17$  &  $1.60 \pm 0.13$  &  $<8.57$  &  \textcolor{black}{$<20.78 $}  \\ \hline \\ 


SN 2021sjt  &  Keck1+LRIS/\textcolor{black}{110}  &  LCOGT1m/\textcolor{black}{110}  &  $19.39 \pm 0.20$  &  $10.13 \pm 0.86$  &  $1.15 \pm 0.08$  &  \textcolor{black}{$2.68 - 8.42$}  \\ \hline \\ 


\end{tabular}  \label{table_neb} 
\end{center} 
\end{table*}

\subsubsection{Modeling the shock cooling powered emission peak}
\label{section:lcfit}

\begin{table*} 
\begin{center} 
\caption{Summary of the best-fit parameters obtained from shock cooling modeling of the first peak.} 
\begin{tabular}{ccccc} 
\hline  \\  

ZTF Source & $E_\mathrm{ext}$ & $R_\mathrm{ext}$ & $M_\mathrm{ext} $ & $t_\mathrm{exp}$  \\    & ($10^{50} \mathrm{erg}$)   & ($\mathrm{R_{\odot}}$)  & ($10^{-2}$ \msun)   & (JD)\\\\ \hline  \\ 

SN 2019ehk/ZTF19aatesgp  &  $3.95^{+0.34}_{-0.32}$  &  $108^{+10}_{-9}$    &  $20.26^{+0.99}_{-0.96}$    &  $2458602.08^{+0.01}_{-0.01}$   \\ \hline \\ 

SN 2021M/ZTF21aaabwfu  &  $7.13^{+1.77}_{-1.91}$  &  $34^{+11}_{-6}$    &  $32.38^{+3.90}_{-4.66}$    &  $2459213.53^{+0.05}_{-0.02}$   \\ \hline \\ 

SN 2020sbw/ZTF20abwzqzo  &  $0.15^{+0.06}_{-0.03}$  &  $2938^{+1103}_{-1334}$    &  $1.68^{+0.34}_{-0.16}$    &  $2459086.57^{+0.07}_{-0.08}$   \\ \hline \\ 

SN 2021pb/ZTF21aabxjqr  &  $4.57^{+0.84}_{-0.71}$  &  $118^{+19}_{-16}$    &  $30.97^{+2.63}_{-2.45}$    &  $2459218.01^{+0.02}_{-0.01}$   \\ \hline \\ 

SN 2019pof/ZTF19abxtcio  &  $0.10^{+0.06}_{-0.03}$  &  $368^{+188}_{-149}$    &  $8.52^{+1.77}_{-1.33}$    &  $2458730.35^{+0.48}_{-0.25}$   \\ \hline \\ 

SN 2018gjx/ZTF18abwkrbl  &  $14.08^{+0.40}_{-0.41}$  &  $47^{+1}_{-1}$    &  $41.00^{+0.86}_{-0.88}$    &  $2458376.34^{+0.01}_{-0.01}$   \\ \hline \\ 

SN 2019hvg/ZTF19abacxod  &  $1.85^{+0.34}_{-0.30}$  &  $121^{+22}_{-18}$    &  $23.36^{+2.05}_{-1.97}$    &  $2458651.03^{+0.04}_{-0.02}$   \\ \hline \\ 

SN 2018jak/ZTF18acqxyiq  &  $9.92^{+0.06}_{-0.12}$  &  $234^{+7}_{-7}$    &  $54.19^{+1.19}_{-1.15}$    &  $2458442.50^{+0.01}_{-0.01}$   \\ \hline \\ 

SN 2021sjt/ZTF21abjyiiw  &  $0.07^{+0.01}_{-0.01}$  &  $1946^{+805}_{-504}$    &  $2.47^{+0.26}_{-0.26}$    &  $2459400.70^{+0.03}_{-0.01}$   \\ \hline \\ 

\end{tabular}  \label{table_piro} 
\end{center} 
\end{table*}

\begin{table*} 
\begin{center} 
\caption{Summary of the best-fit physical parameters obtained by radioactive decay modeling of the second peak of the SNe in our sample. The `Ar' subscript refers to the \citet{Arnett89} model while the `KK' subscript refers to the \citet{KhatamiKasen2019} model.} 
\begin{tabular}{cccccccc} 
\hline  \\  

Source & $M_\mathrm{Ni-Ar}$\footnote{Lower limit is calculated by assuming no host extinction.}  & $M_\mathrm{Ni-KK}$ & $M_\mathrm{ej-Ar}$ & $M_\mathrm{ej-KK}$  & $E_\mathrm{kin}$   & $t_0$   \\  & (\msun)  & (\msun)  & (\msun)  & (\msun)  & ($10^{51} \mathrm{erg}$) & (days)   \\\\ \hline  \\ 

SN 2019ehk/ZTF19aatesgp  &  $ 0.02-0.08$    &  $0.08$    &  $0.35^{+0.01}_{-0.01}$    &  $0.32$    &  $0.21^{+0.01}_{-0.01}$    &  $37.79^{+0.98}_{-0.96}$   \\ \hline \\ 

SN 2021M/ZTF21aaabwfu  &  $0.04-0.13$    &  $0.14$    &  $1.05^{+0.64}_{-0.54}$    &  $0.79$    &  $0.76^{+0.46}_{-0.39}$    &  $52.83^{+2.86}_{-2.76}$   \\ \hline \\ 

SN 2020sbw/ZTF20abwzqzo  &  $0.02-0.13$    &  $0.14$    &  $0.77^{+0.09}_{-0.07}$    &  $0.57$    &  $0.29^{+0.04}_{-0.03}$    &  $59.76^{+1.57}_{-1.60}$   \\ \hline \\ 

SN 2021pb/ZTF21aabxjqr  &  $0.10- 0.12$    &  $0.13$    &  $3.57^{+2.74}_{-0.58}$    &  $2.21$    &  $5.46^{+4.19}_{-0.89}$    &  $56.84^{+29.20}_{-29.00}$   \\ \hline \\ 

SN 2019pof/ZTF19abxtcio  &  $0.01-0.03$    &  $0.04$    &  $0.77^{+0.37}_{-0.14}$    &  $0.59$    &  $0.46^{+0.22}_{-0.08}$    &  $40.17^{+1.39}_{-1.44}$   \\ \hline \\ 

SN 2018gjx/ZTF18abwkrbl  &  $0.04- 0.05$    &  $0.05$    &  $0.20^{+0.02}_{-0.02}$    &  $0.19$    &  $0.08^{+0.00}_{-0.00}$    &  $59.32^{+1.18}_{-1.13}$   \\ \hline \\ 

SN 2019hvg/ZTF19abacxod  &  $0.02-0.13$    &  $0.15$    &  $1.15^{+0.08}_{-0.07}$    &  $0.81$    &  $0.56^{+0.04}_{-0.03}$    &  $47.30^{+3.76}_{-2.90}$   \\ \hline \\ 

SN 2018jak/ZTF18acqxyiq  &  $0.03-0.12$    &  $0.11$    &  $0.72^{+0.04}_{-0.04}$    &  $0.62$    &  $0.73^{+0.04}_{-0.04}$    &  $93.83^{+4.40}_{-6.61}$   \\ \hline \\ 

SN 2021sjt/ZTF21abjyiiw  &  $0.01-0.05$    &  $0.05$    &  $1.15^{+0.03}_{-0.03}$    &  $0.83$    &  $0.83^{+0.02}_{-0.02}$    &  $50.23^{+0.77}_{-0.74}$   \\ \hline \\ 

\end{tabular}  \label{table_arnett} 
\end{center} 
\end{table*}

All the sources in our sample have a double-peaked lightcurve, which is usually seen in Type IIb SNe. The timescale of the first peak, the initial blue color and high temperatures indicate that the first peak is dominated by cooling emissions from the shock heated extended envelope \citep{Rabinak2011, nakarpiro2014, piro2015}. We note that some of the SNe from our sample do not have well sampled first peaks on both the rising and fading parts.  We use the model by
\citet{Piro2021} to fit for the shock cooling breakout peak. This model constrains the extended material radius ($R_\mathrm{ext}$), mass ($M_\mathrm{ext}$), energy ($E_\mathrm{ext}$) and the explosion time ($t_\mathrm{exp}$). We first select the appropriate epochs of
the lightcurve peaks to model as shock cooling and then
we fit the multi-band data using the Python \emcee\ package. The best-fit values and fits for each source are provided in Table \ref{table_piro} and Appendix \ref{appendix:piro}. 


\subsubsection{Modeling the radioactively powered lightcurve peak}
\begin{figure}[ht!]
    \centering
    \includegraphics[width=1\columnwidth]{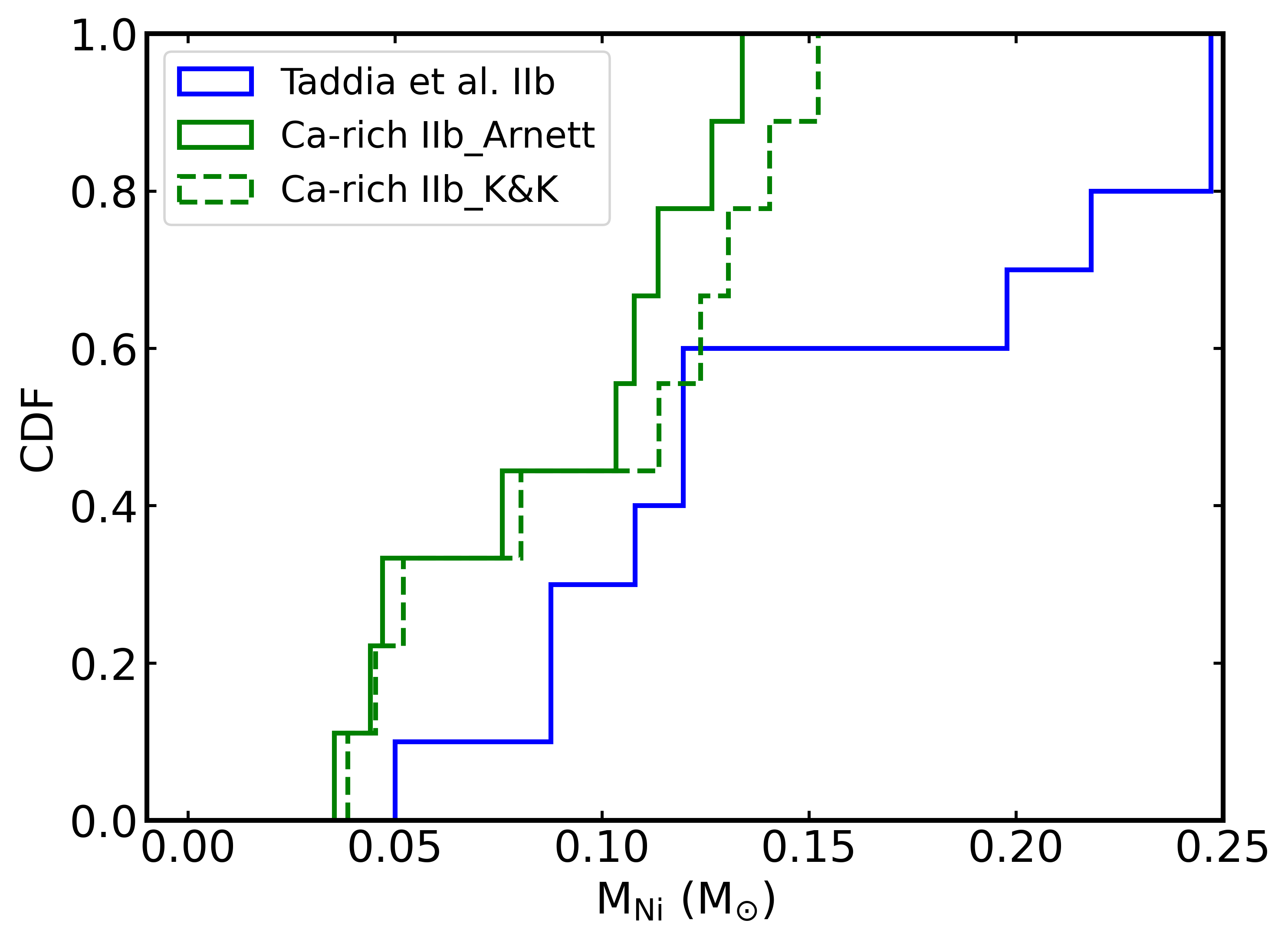}
    \caption{\Nif-mass distribution for our sample of SNe IIb compared to the SN IIb population presented in \citet{Taddia2018}.}
    \label{fig:knmodelcomp}
\end{figure}

\begin{figure}[ht!]
    \centering
    \includegraphics[width=1\columnwidth]{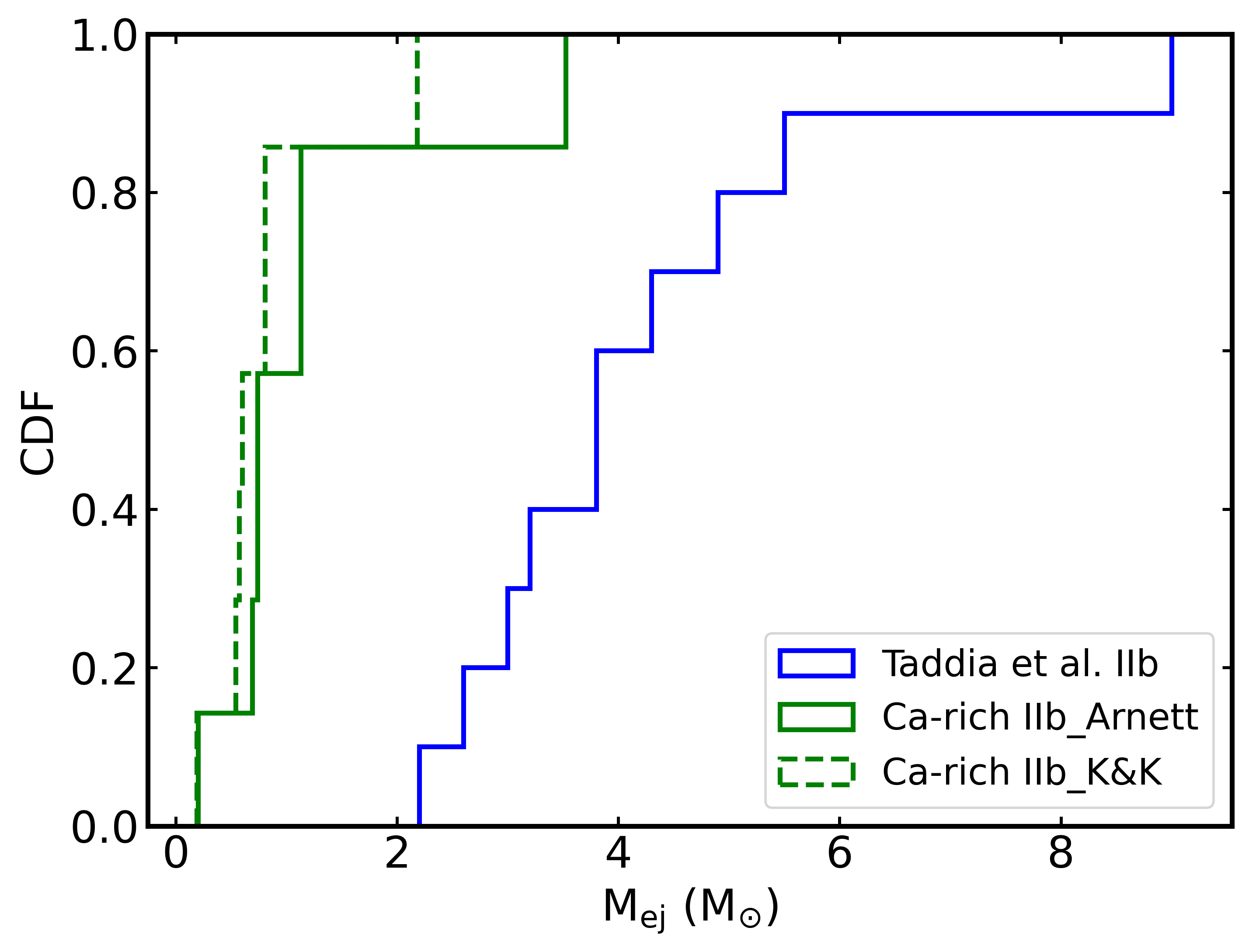}
    \caption{Ejecta mass distribution for our sample of SNe IIb compared to the SN IIb population presented in \citet{Taddia2018}.}
    \label{fig:ejmodelcomp}
\end{figure}

\begin{figure}[ht!]
    \centering
    \includegraphics[width=1\columnwidth]{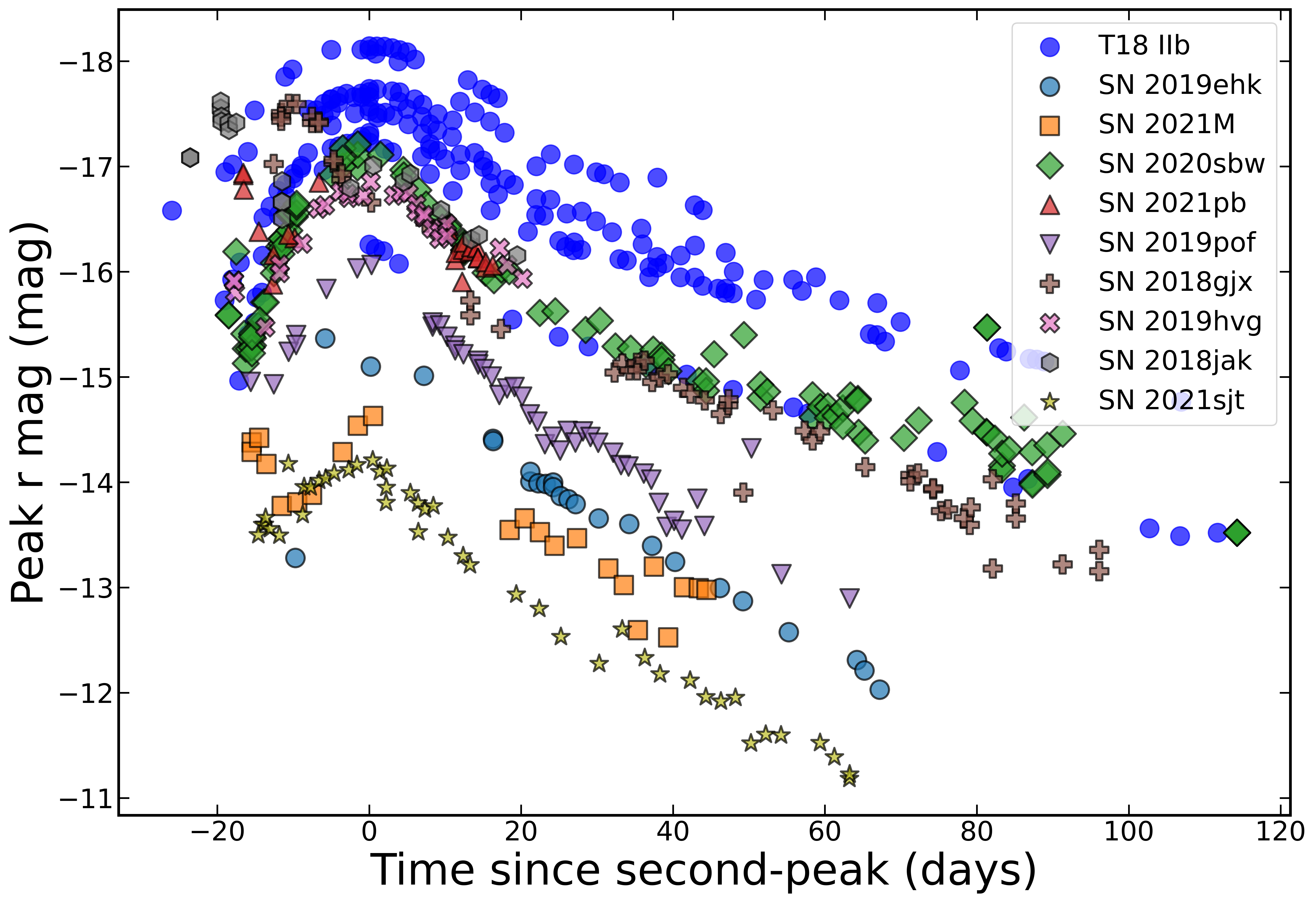}
    \caption{\textcolor{black}{Comparison of the lightcurves of the Ca-rich Type IIb SNe in our sample (in filled shapes of different types) with canonical Type IIb SNe lightcurves from \citet{Taddia2018} (in blue circles). The y-axis shows the absolute magnitude (mag), with galactic extinction correction. The x-axis shows the number of rest-frame days since the second peak.}}
    \label{fig:T19lc}
\end{figure}

\begin{figure}[ht]
    \centering
    \includegraphics[width=1\columnwidth]{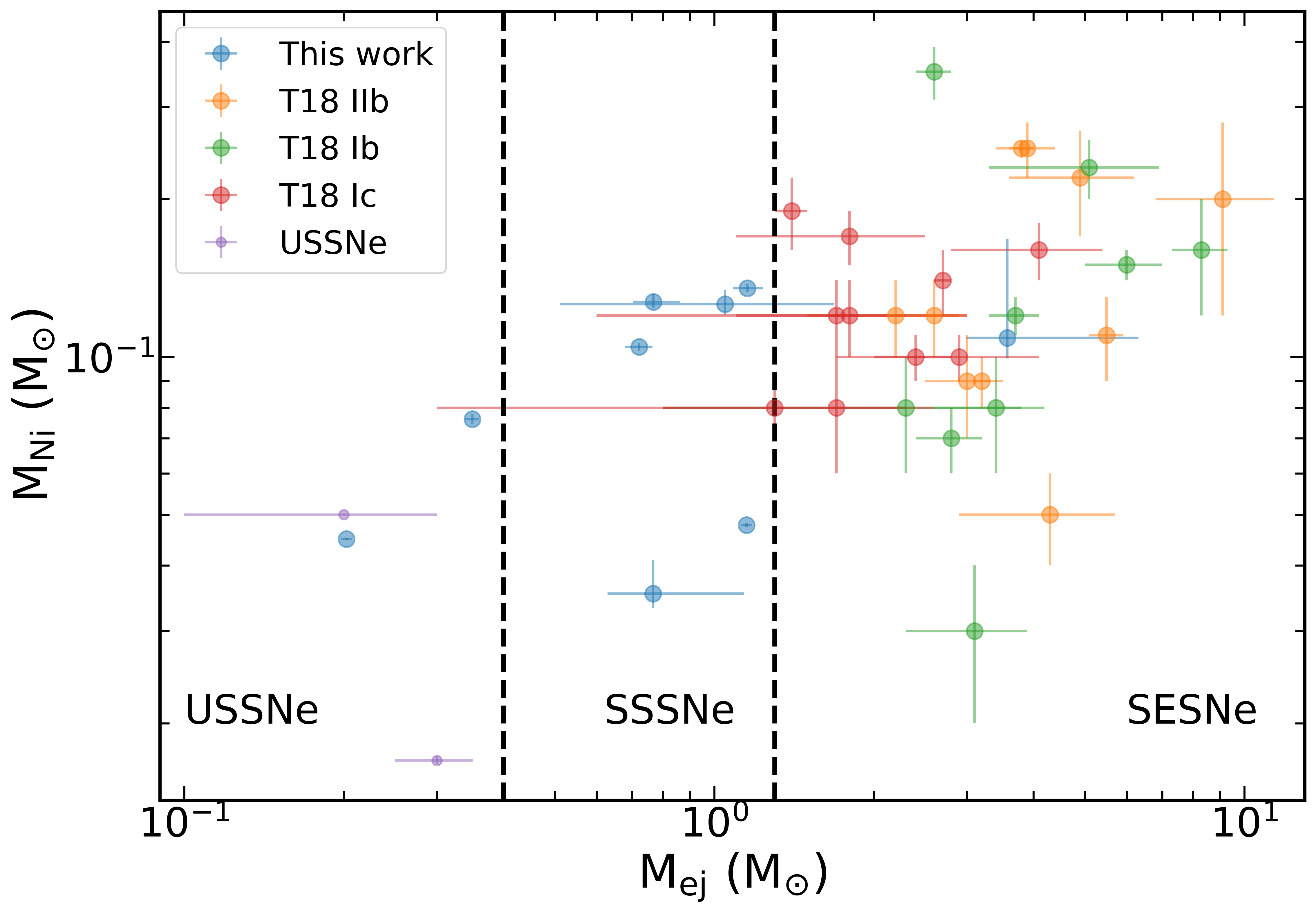}
    \caption{Comparison of the \Ni-mass and the ejecta mass \textcolor{black}{based on the Arnett fit} for our sample and usual stripped-envelope Type IIb SNe from \citet{Taddia2018} and USSNe from \citet{De2018a}, \citet{Yao2020}. We find that the SSSNe in our sample is a transient class between the USSNe and the usual stripped-envelope SNe (SESNe) (rough boundaries indicated by the vertical dashed lined). We also note that there is a region in the parameter-space which is unoccupied by the SESN \textcolor{black}{population}.}
    \label{fig:mnimej2}
\end{figure}

After fitting the initial cooling peaks in our lightcurves, we use the best-fit parameters to estimate the contribution of the cooling emission
to the bolometric lightcurve of each source. We then subtract the
cooling component from the bolometric lightcurves obtained
from blackbody fitting. The subtracted lightcurves now correspond to that of the radioactively powered peak. We fit for this peak using two methods. First, we use the analytical model described in \citet{Arnett89}
and \citet{Valenti2008}. We attempt to constrain
the nickel mass (\Mni), characteristic photon diffusion
timescale ($\tau_m$) and characteristic 
 $\gamma$-ray diffusion timescale
($t_o$). We use relations from \citet{Wheeler2015} that give the
ejecta mass (\Mej) and kinetic energy in the ejecta ($E_\mathrm{kin}$) as
a function of ($\tau_m$) and photospheric velocity ($v_\mathrm{ph}$).
We obtain $v_\mathrm{ph}$ measured using the  He I line from the photospheric spectra closest to the peak epoch as described in Section \ref{sec:photspectra}. We also use the lightcurve analytical models from \citet{KhatamiKasen2019} to derive the explosion parameters. Further details on the model fitting can be found in Appendix B9 in \citet{Yao2020}. Table \ref{table_arnett} lists the best-fit parameters. Figures \ref{fig:knmodelcomp} and \ref{fig:ejmodelcomp} show the comparison of the cumulative distribution functions
for \Mni\ and \Mej\ of our sample and the \citet{Taddia2018} sample of Type IIb SNe. A two-sided K-S test of the ejecta mass distribution gives a p-value $<$ 0.001, thus implying that our sample contains SNe with lower ejecta mass compared to the literature sample of SNe IIb.
Our sample indicates that there is an empty phase space with low ejecta mass but high nickel mass and high ejecta mass but low nickel mass (see Figure \ref{fig:mnimej2}). Further analysis on this with a larger sample of Type IIb SNe is left for future work.
We plot the lightcurves of the Ca-rich Type IIb SNe sample along with canonical Type IIb SNe from \citet{Taddia2018} in Figure \ref{fig:T19lc}.


\subsection{Host Galaxy Spectral Energy Distribution Modeling} \label{sec:hostgal}

\begin{figure}
\centering
\includegraphics[width=1\columnwidth]{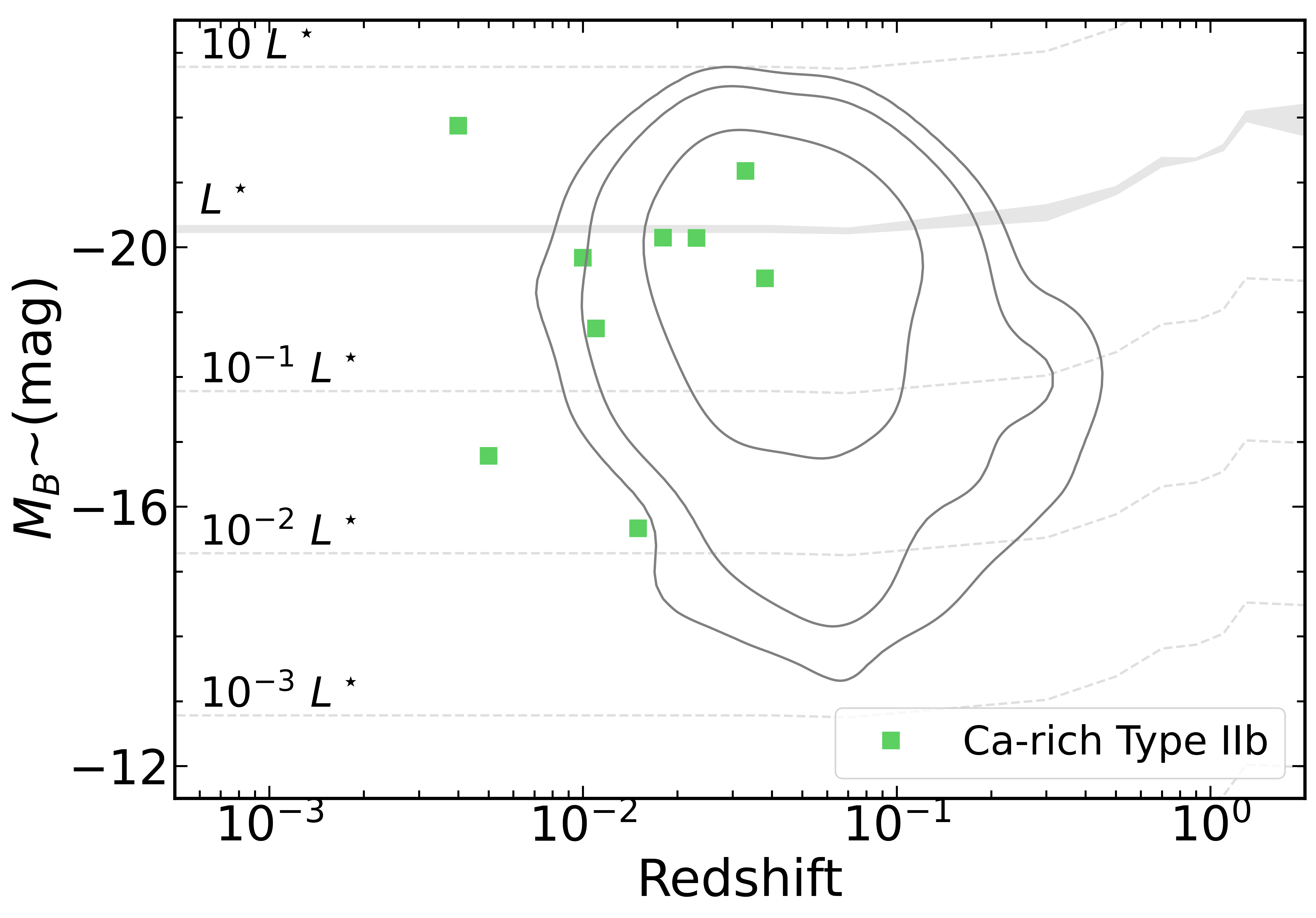}
\caption{The absolute $B$-band magnitude of CCSN host galaxies as a function of redshift. Our hosts are found in the least-luminous ($10^{-3}~L^\star$) to the most luminous star-forming galaxies ($\lesssim5~L^\star$) ($L^\star$ is the characteristic luminosity of the $B$-band luminosity function of star-forming galaxies). Most Ca-rich IIb hosts have luminosities of $10^{-2}$ to a few $L^\star$, similar to regular CCSNe \citep[indicated by the contours encircling 68, 90 and 95\% of the PTF+iPTF CCSN sample][] {Schulze2020a}. We also indicate the $L^\star$ presented in \citet{Faber2007} and multiples of it with gray.
}
\label{fig:host_MB}
\end{figure}

\begin{figure}
\centering
\includegraphics[width=1\columnwidth]{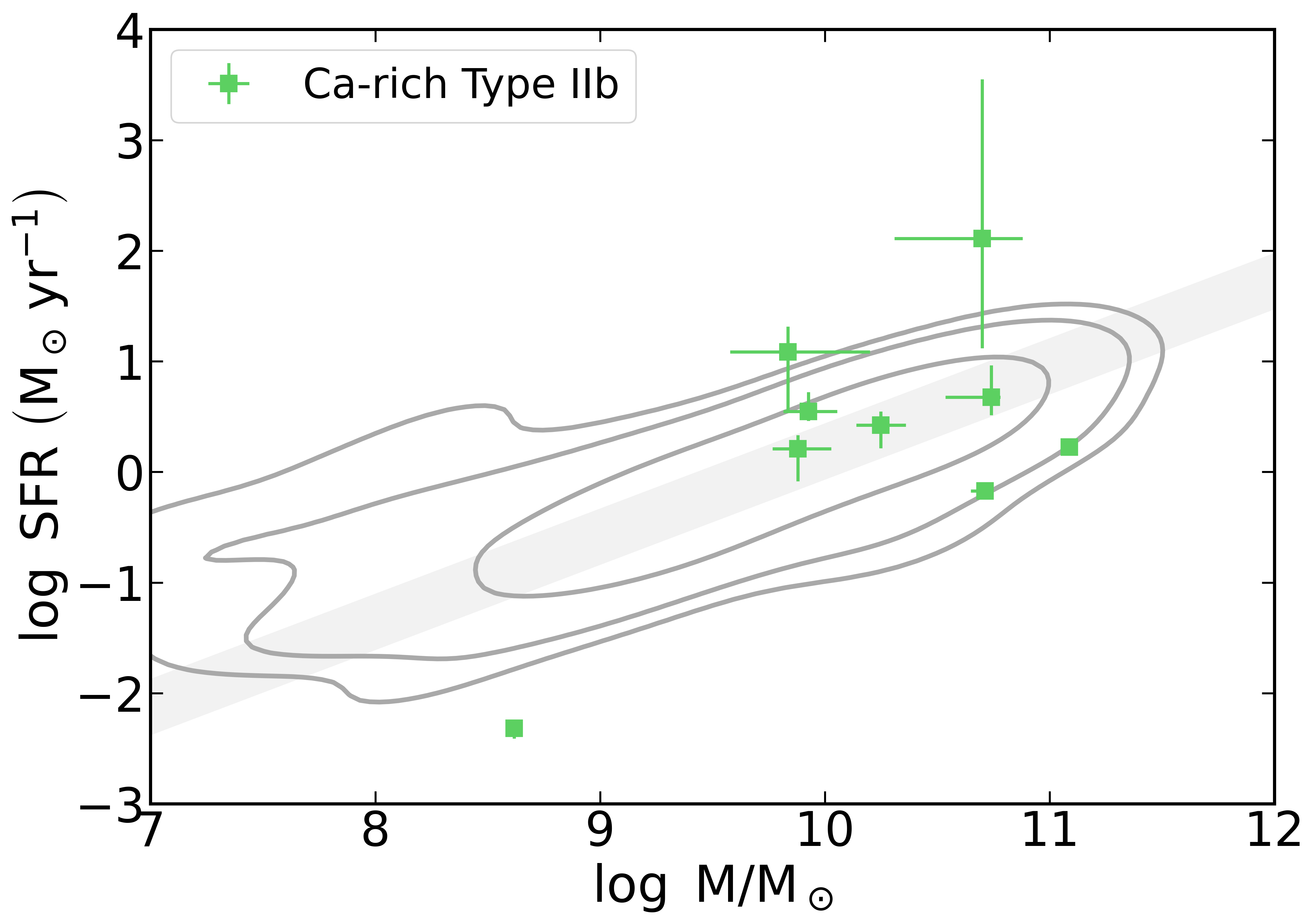}
\caption{The host population in the mass-SFR plane. All sources exploded in star-forming galaxies. This is illustrated by their location with respect to the main sequence of star-forming galaxies (grey-shaded region). The majority of hosts also have properties consistent with those of CCSNe from the PTF+iPTF surveys \citep[grey contours indicate the region encircling 68, 90 and 95\% of the sample][]{Schulze2020a}}.
\label{fig:sfr_mass}
\end{figure}

\begin{figure}
\centering
\includegraphics[width=1\columnwidth]{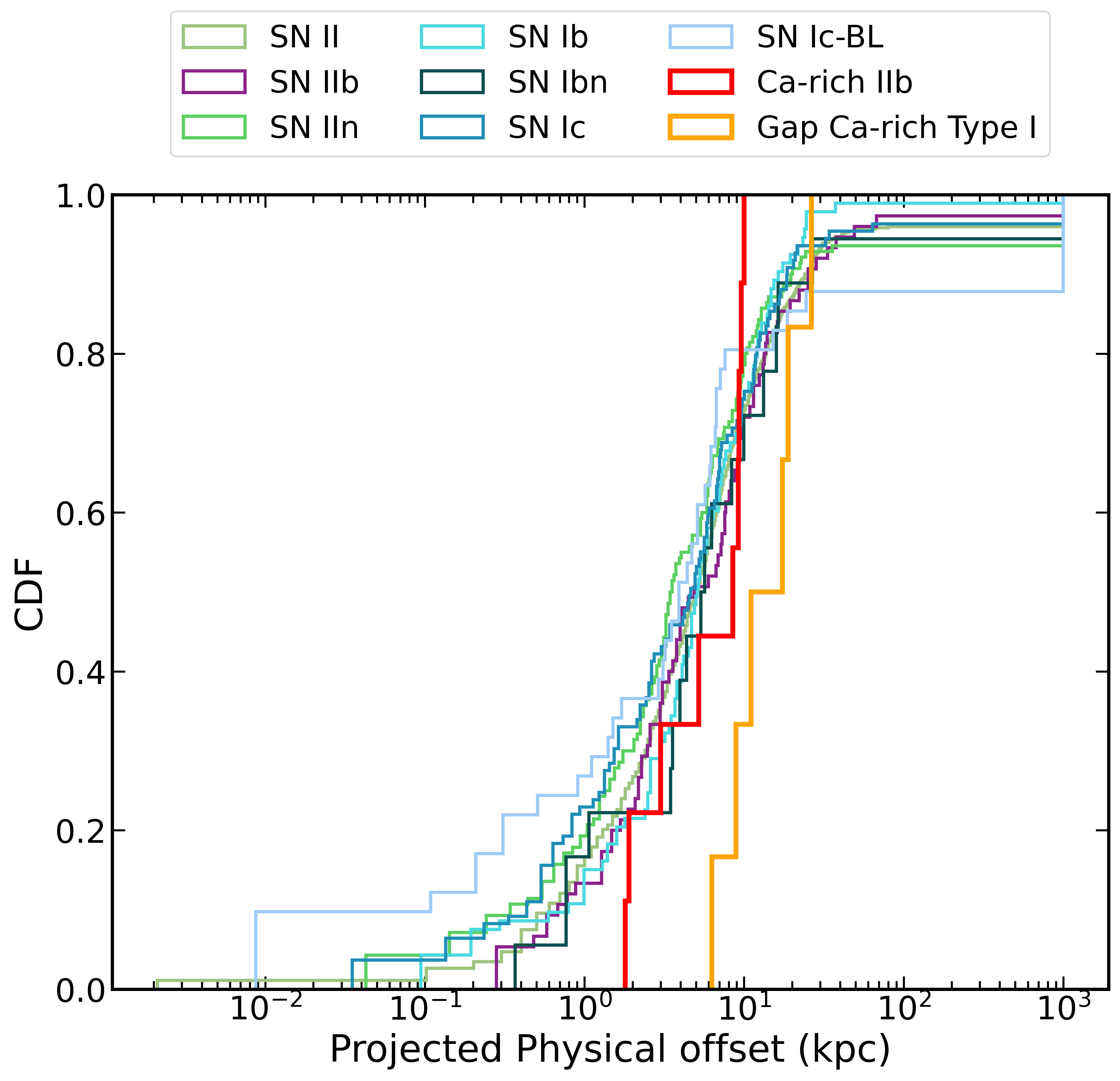}
\caption{The projected physical offset distribution of all the SNe in our sample compared with the distribution for other types of SNe measured in \citet{De2020}. We note that there is a distinct offset from the Gap Ca-rich Type I SNe, known to occur in quenched environments.}
\label{fig:hostoffset}
\end{figure}

We model the spectral energy distributions (SEDs) of the host galaxies with the software package \cigale{} \citep[Code Investigating GALaxy Evolution;][] {Burgarella2005, Noll2009,  Boquien2019}. We use the \citet{Chabrier2003} initial mass function and the \citet{Bruzual2003} simple stellar population model to compute the stellar emission. Furthermore, we assume a star-formation history with the following functional form: $t \times \exp\left(-t/\tau\right)$, where $t$ is the age of the SFH episode and $\tau$ is the $e$-folding timescale. We fix the \cigale{} ionization parameter log $\text{U}_{\text{ion}}$ as $-2$. This is used to calculate the nebular emission from the ionized gas in \ion{H}{2} regions. We adopt a modified \citet{Calzetti2000} attenuation curve to model the dust attenuation. We include dust emission based on the \citet{Dale2014} dust templates. Further details on the different models can be found in \citet{Boquien2019}. We generated 24385536 models based on the mentioned configuration, from which we chose the best fit using Bayesian inference. Table \ref{tab:cigalemod} lists the parameters used to generate the models and Table \ref{tab:galaxyprop} lists the best-fit parameters for all the host galaxies.

The hosts of all transients were detected in images of the SDSS, Pan-STARRS, and the DESI Imaging Legacy Surveys. Figure \ref{fig:host_MB} shows the $B$-band luminosities of the hosts for our sample. The entire distribution spans a wide range from $M_B\approx-15.7$~mag to $M_B\approx-21.8$~mag. This interval covers the range from $10^{-2}\, L_\star$ to $10\, L_\star$, where $\, L_\star$ is the knee of the $B$-band luminosity functions reported in \citet{Ilbert2005a}, covering the entire parameter space of star-forming galaxies. The majority of the hosts are found in galaxies between a few $10^{-2}$ and a few $L^\star$. These host properties are identical to hosts of usual CCSNe. This is illustrated by the contours encircling 68, 90 and 95\% of the PTF+iPTF CCSN sample \citep{Schulze2020a} that includes 888 objects from all major CCSN classes.

To compare the physical properties of the host galaxies, we present them in the mass-SFR plane (Fig. \ref{fig:sfr_mass}). The hosts are located in a narrow region of the parameter space that is occupied by a general population of star-forming galaxies, called the main sequence of star-forming galaxies (indicated by the grey shaded region; \citealt{Elbaz2007a}). A small minority  of our SNe occurred in galaxies that lie clearly above the galaxy main sequence and are experiencing a starburst.  Like in Figure \ref{fig:host_MB}, we overlay the 68, 90 and 95\% contours of the PTF CCSN host sample. The hosts of regular CCSNe occupy exactly the same parameter space, including the starburst regime \citep[e.g.,][]{Taggart2021a}. Most of the host galaxies are actively star-forming, which shows that the core-collapse Ca-rich sample is a different class from the usual Ca-rich sample, which are usually found in the outskirts of quenched galaxies. This is also evident from Figure \ref{fig:hostoffset}, which shows that the projected physical offset distance for Gap Ca-rich thermonuclear transients is systematically higher \textcolor{black}{(p-value = 0.06 for a two-sided K-S test)} than that of our Ca-rich Type IIb sample and the general core-collapse population \citep[for the sample presented in][]{De2020}.

\section{Discussion} 
\label{sec:discussion}

In this paper, we analyze the photometric and spectroscopic properties of the sample of a distinct class of Ca-rich Type IIb SNe. The sources in our sample are all core-collapse and not thermonuclear in origin. This is evident from the presence of photospheric hydrogen in the spectra (see Section \ref{sec:photspectra}) which suggests $M_{H} \gtrsim 0.02-0.03\ M_\odot$ \citep{Hachinger2012}. The usual Ca-rich Type I events involve CO + He binary WDs progenitors, that are expected to be very deficient in hydrogen \citep[$M_{H}$ $< 10^{-4} M_\odot$;][]{Podsiadlowski2003, Zenati2019}. Hence, the formation pathway of our sample is distinct from the thermonuclear Ca-rich gap transients found in old environments. The host galaxy properties described in Section \ref{sec:hostgal} are also consistent with that of the core-collapse population exploding in star-forming galaxies. The observed photospheric velocities, which are consistent with canonical stripped-envelop SNe, also rules out the fallback SNe scenario which have an expected velocity of $\sim$ 3000 $\mathrm{km\ s^{-1}}$ \citep{Moriya2010}. In this section, we discuss the constraints on the progenitors and implications on the progenitor channels for this class.


\subsection{Constraints on progenitor mass}
\subsubsection{Constraining progenitor mass using nebular spectra} \label{nebulara}

\begin{figure}
    \centering
    \includegraphics[width=1\columnwidth]{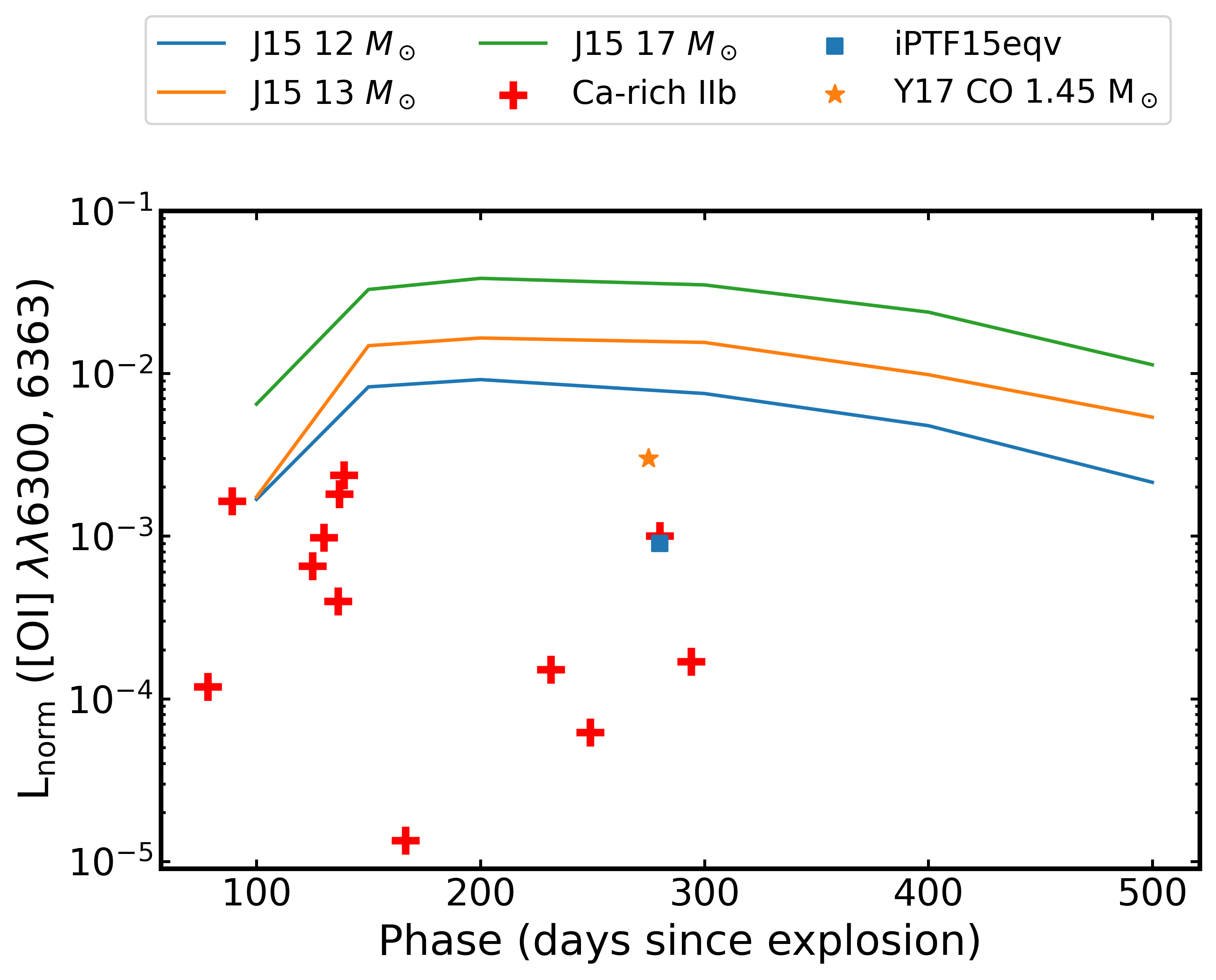}
    \caption{The red crosses shows the [O~I] luminosities calculated from the nebular spectra for the Ca-rich Type IIb sample. The luminosity is normalized to the \Cif\ decay energy. We compare these to models of stripped envelope core-collapse SNe from \citet{Jerkstrand2015}. We also show the measured normalized [O~I] luminosities from the nucleosynthesis calculations of \citet{Yoshida2017} and the Ca-rich IIb iPTF 15eqv.}
    \label{fig:olum}
\end{figure}

\begin{figure}
\centering
\includegraphics[width=1\columnwidth]{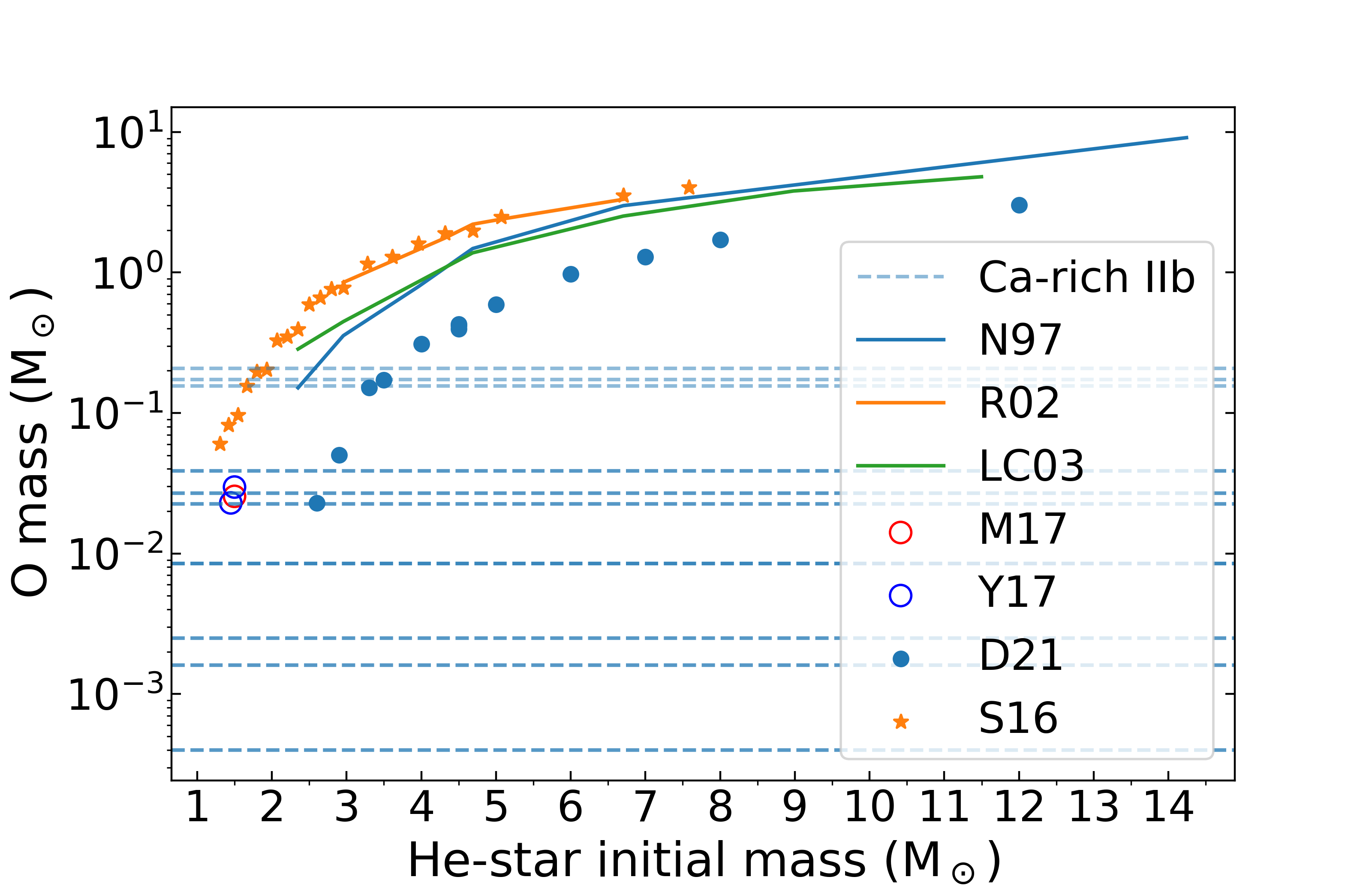}
\caption{The measured O mass \textcolor{black}{lower limits} are shown with the horizontal dashed lines. The O yield for the ejecta arising from the He-star progenitor models, which assume an evolution in a binary system  \citep{Dessart2021} are shown in solid blue dots and those arising from the single-star models  \citep{Sukhbold2016} are shown in filled orange stars. We also plot the O mass for nucleosynthetic models of higher ZAMS stars \citep{Nomoto1997N, Rauscher2002, Limongi2003} and for lower ZAMS stars from USSNe models of \citet{Yoshida2017} and \citet{Moriya2017}. }
\label{fig:Omass_dessart}
\end{figure}

For all the SNe in our sample, we have at least one nebular spectrum obtained using the Low Resolution Imaging Spectrograph (LRIS) on the Keck I telescope. We calculate the [Ca II]/[O I] flux ratio and the [O I] luminosity using the procedure described in Section \ref{section:nebspectra} (see Table \ref{table_neb}). Next, we use the measured [O~I] luminosity to calculate the oxygen abundance and subsequently the progenitor mass. We use the following analytical relation provided in \citet{Uomoto1986} to calculate the minimum O mass required for a given [O~I] luminosity:
\textcolor{black}{$M_\mathrm{O} = 10^8\ (\mathrm{L}[\mathrm{O~I}]/4\pi \mathrm{D^2_{cm}})\ \mathrm{D^2_{Mpc}}\ \mathrm{exp}(2.28/\mathrm{T}(10^4 \mathrm{K}))$ \Msun, where $\mathrm{L}[\mathrm{O~I}]$ is the luminosity in erg/s of [O I] $\lambda \lambda$6300, 6364}. The analytical formula is valid in the high density limit where the electron density is above the [O~I] critical density ($\sim 7 \times 10^5 \mathrm{cm^{-3}}$), and is estimated to hold in our case for estimated ejecta mass in the range of $0.5-2$ \msun. We use typical temperature values estimated from the [O~I] emission in other core-collapse SNe of $\approx$ $3500-4000$ K \citep{Sollerman1998, Elmhamdi2011} to get lower limits on the O mass in the range of $\approx$ $0.005-0.1$ \Msun. The lower and upper limits of the O mass provided in Table \ref{table_neb} correspond to 4000 K and 3400 K, respectively.

We compare the observed [O~I] luminosity and measured oxygen mass \textcolor{black}{lower limits} with theoretical models to constrain the progenitor mass. \citet{Jerkstrand2015} presented the dependence of [O~I] luminosity with progenitor mass using detailed nebular phase models of stripped-envelope SNe (see Figure \ref{fig:olum}). We see that the nebular models of progenitors with mass $> 12\ \Msun$ consistently overestimate the [O~I] luminosity. This suggests a much lower progenitor core mass for our sample. 
In Figure \ref{fig:Omass_dessart}, we plot model tracks showing the steep dependence of O mass on the initial ZAMS mass obtained from canonical core-collapse nebular models \citep[taken from][]{Nomoto1997N, Rauscher2002, Limongi2003}. We consistently find that the ZAMS mass required to explain the observed O mass \textcolor{black}{lower limits} of our sample can \textcolor{black}{be} less than $ 12\ \Msun$.


\citet{Dessart2021} presented a set of 1D non-local thermodynamic equilibrium radiative transfer calculations for nebular-phase stripped SNe. In models from He stars with an initial mass $>$6 \Msun, they find that the [O~I] line strength is of comparable or greater strength than [Ca~II] lines. The initial helium star mass reflects the mass of the helium core at the time of central helium ignition. In contrast, they show that models from lower mass He stars exhibit a weak [O~I], strong [Ca~II]. The high [Ca~II]/[O~I] flux ratio we observe for our sample is thus consistent with low initial He-star mass. Figure \ref{fig:Omass_dessart} compares the measured O mass lower limits for our sample with the synthesized O mass from the He-star progenitor models which assume an evolution in a binary system  \citep{Dessart2021} and with those arising from single-star models \citep{Sukhbold2016}. In both the binary and single star systems, core collapse is reached with a residual H-rich envelope. Again, we find that the observed oxygen yield is consistent with the lowest ZAMS mass ($< 12\ \Msun$) models for stripped He stars with initial mass less than $\sim$ 3 \Msun. We use the relations in \citet{Woosley2015} to get the ZAMS mass from the initial He-star mass.  

To compare with such low ZAMS masses, we use estimates of the O synthesized for the case of the highly stripped He cores of ultra-stripped SNe. These are modeled as relatively low-mass He stars ($< 3.5\ M_\odot$), that are stripped down by a close binary companion, leaving behind CO cores of $\approx$ 1.45$-$1.6 \msun\ at the time of explosion \citep{Tauris2015}. We note that the CO core mass is a good tracer of the progenitor ZAMS as it is insensitive to the binary stripping that occurs in the very late stages of stellar evolution \citep{Fransson1989, Jerkstrand2014, Jerkstrand2015}. We find that the O-yields for such strongly-stripped stars with low CO-core mass are comparable to the low O-yield of sources in our sample (see Figures \ref{fig:olum} and \ref{fig:Omass_dessart}).


We caution that we assume that the radioactive power deposited in the O rich shells of the ejecta is released via cooling in the [O~I] lines. We note that this is dependent on the presence of impurity species, e.g., \citet{Dessart2020} showed that even a small amount of Ca mixed into O-rich regions can dramatically weaken the cooling by [O~I]. However, it has been shown that mixing is not significant for core-collapse SNe. Detailed modeling of core-collapse SNe has shown that [OI] emission arises from the outer O-rich layers produced during the hydrostatic burning phase,  while the [Ca II] line is the primary coolant in the Si-rich layers \citep{Jerkstrand2015, Dessart2020}. Also, \cite{Polin2021} show that pollution of $^{40}$Ca even at the 1\% level will cause a nebular region to completely cool through [Ca~II] emission suggesting that if these ejecta regions were mixed it would be difficult to see any contribution from [O~I]. 

\subsubsection{Constraining progenitor mass using lightcurves}
\label{section:dis_lc}

We fit the double-peaked lightcurves of the SNe sample using the \citet{Piro2021} and \cite{Arnett89} models as described in Section \ref{section:lcfit}. The best fit parameters of the shock-cooling breakout peak and the radioactive-powered peak such as $M_\mathrm{Ni}, M_\mathrm{ej}, E_\mathrm{kin}, R_\mathrm{env}$ are tabulated in Tables \ref{table_piro} and \ref{table_arnett}.

\begin{figure}
\centering
\includegraphics[width=1\columnwidth]{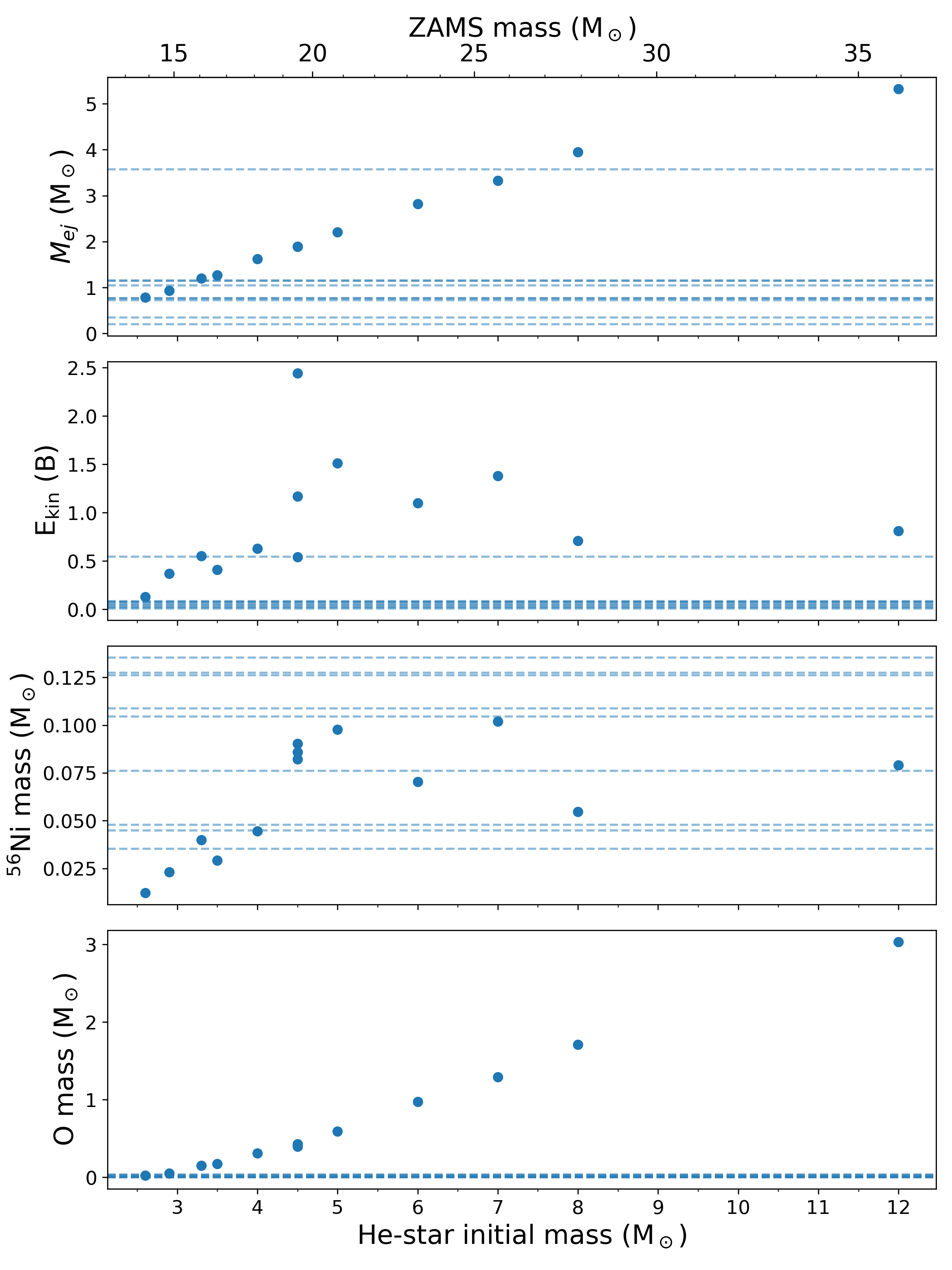}
\caption{Comparison of the ejecta properties we observe (horizontal dashed lines) for our sample with that for the explosion models based on He stars evolved with the nominal mass-loss rate from \citet{Dessart2021} (solid dots). From top to bottom, we show the ejecta mass $M_\mathrm{ej}$, ejecta kinetic energy $E_\mathrm{{kin}}$, the \Ni-mass, and the oxygen yield.}
\label{fig:dessart_compare}
\end{figure}

In order to constrain the progenitor properties, we plot the evolutionary track of the predicted ejecta properties as a function of He-star initial mass or initial ZAMS mass from \citet{Dessart2021}. They use He-star models of \citet{Woosley2019} with initial He-star masses spanning the  range  from  2.6 \Msun\ to  13 \Msun\  that were evolved to the onset of core collapse. The ejecta properties shown by the solid dots in Figure \ref{fig:dessart_compare} are obtained after performing nebular-phase NLTE steady-state and time-dependent radiative transfer calculations based on prescribed He-star explosion models of \citet{Ertl2020}. Depending  on  the adopted  pre-SN  mass  loss  recipe, they give  rise  to  ejecta  masses between  0.79 \Msun\  and  5.32 \Msun. The ejecta mass values we obtain are consistent with those predicted for the lower end of the He-star mass stars ($<12$ \Msun). They report that the lower mass progenitors with initial He-star mass of $\approx$ 2.6 \msun, calibration to the Crab and realistic 10 \msun\ single star simulations implies an explosion energy of $\sim 10^{51}$ erg. It can be seen from Figure \ref{fig:dessart_compare} that the kinetic energy of our sample is consistently lower than that predicted for ZAMS stars of $\sim 12\ \Msun$. We note that most SNe in our sample have low ejecta mass while having high Ni mass. This could be due to enhanced mass loss rate of the pre SN He star. \citet{Dessart2021} found that there is a systematic increase in the \Ni\ production by a few factors depending on the mass loss rate and the initial mass. The high Ni mass makes an electron-capture core-collapse scenario unlikely as simulations consistently predict an upper limit of $\sim 10^{-2}$ \msun, that are an order of magnitude lower than those observed for our sample.

\begin{figure}
\centering
\includegraphics[width=1\columnwidth]{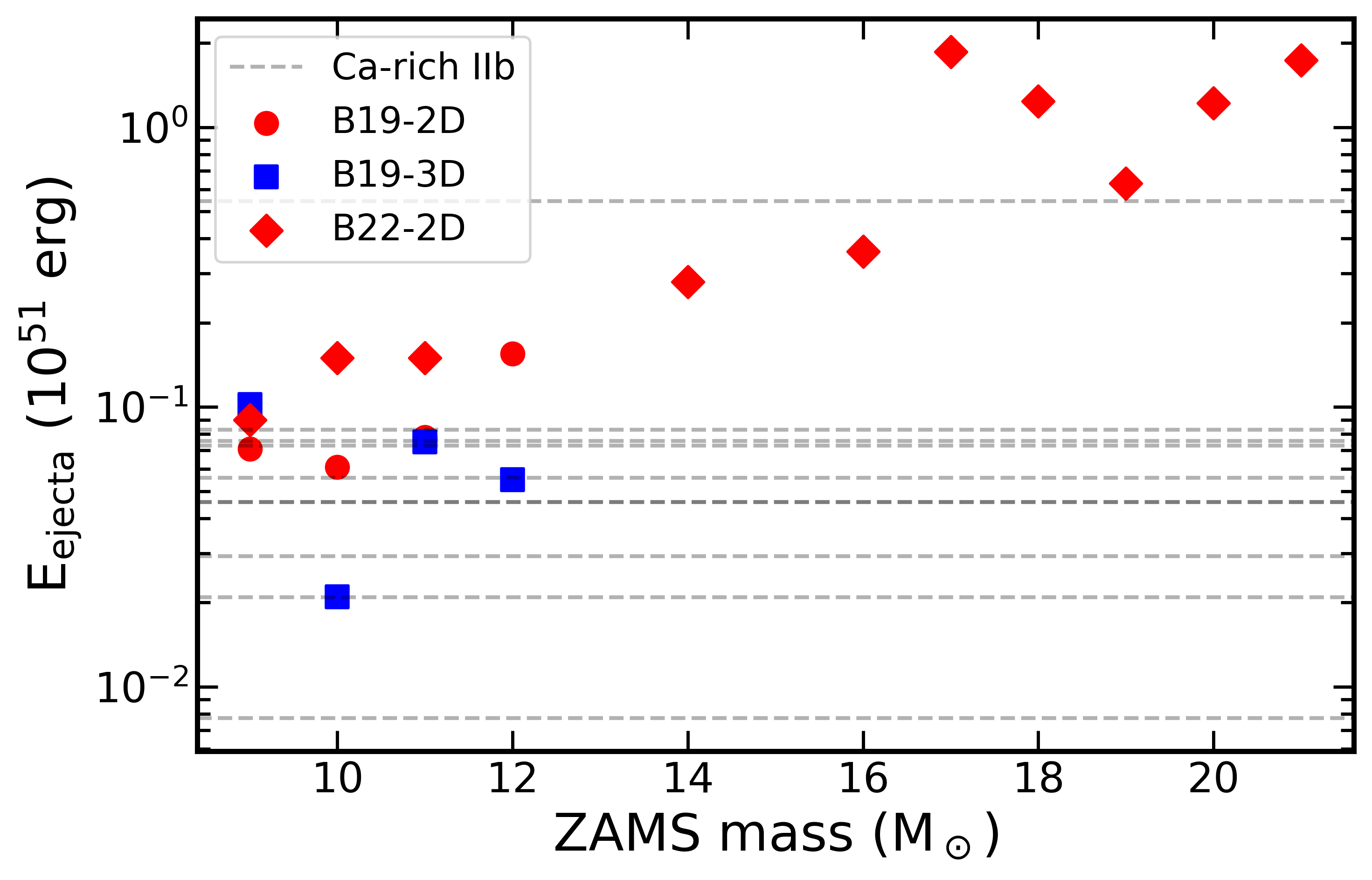}
\caption{Comparison of the observed explosion energy values with predictions from \citet{Burrows2019} (B19), \citet{Burrows2021} (B22). They use the CCSN code \fornax\ to perform 2D and 3D core-collapse supernova simulations for a range of progenitor masses. See Section \ref{section:dis_lc} for details.}
\label{fig:B22comp}
\end{figure}

\textcolor{black}{Next, we compare the observed explosion energy values with predictions from state-of-the-art 2D and 3D CCSN simulations. \citet{Burrows2019} uses the CCSN code \fornax\ to simulate the behaviour of CCSNe at the lower mass end (9, 10, 11, 12, and 13 \Msun) in three-spatial directions while \citet{Burrows2020} conduct nineteen 3D CCSN simulations spanning a wide range of progenitor masses from $9-60$ \msun. Neutrino heating, turbulence behind the stalled shock, energy transfer due to neutrino–matter absorption and neutrino–matter scattering, among other factors, were conducive for the explosion of these low mass progenitors. They consistently find that the lower-mass massive star progenitors in the  9$-$12 \Msun\ range experience lower-energy explosions, while the higher-mass massive stars experience higher-energy explosions. In figure \ref{fig:B22comp}, we compare the predicted energy values from \citet{Burrows2019}, \citet{Burrows2021} with the observed values for our sample.  We find that the explosion energy values ($\sim10^{50}$ erg) we obtain are roughly consistent with the 2D models and completely consistent with the 3D SN simulations for the $\approx 9 - 12$ \Msun\ progenitors. We note that the explosion energy predicted from these models is defined as sum of the gravitational, kinetic, thermal, and recombination energies of the ejecta and thus should be treated as upper limit to the explosion kinetic energy. The models also mention a gap in explodibility for 13 \msun\ in \citet{Burrows2019}, potentially due to \textcolor{black}{shallow} mass density and subtle Si/O interface, which can be explored with a larger sample of well-constrained low mass CCSNe.}

It is has been shown from simulations that progenitors of stripped-envelope SNe with initial masses below 10 \Msun\ expand to large radii (up to 100 \Rsun), while more massive progenitors stay compact \citep[e.g.,][]{Laplace2020}. Only progenitors with low initial masses in the 8$-$9\ \Msun\ range can swell enough for renewed interaction with their companion. This is consistent with the envelope radii of our sample, with values lying in the 100$-$1000 \Rsun\ range. The extent of expansion of a stripped star is critical to understand their fate as the progenitors of core collapse supernovae. The large radii of stripped stars imply that they can potentially fill their Roche lobe again and undergo another phase of mass transfer \citep{Dewi2002, Dewi2003}. Additional phases of mass transfer can produce stars with very low envelope masses, possibly explaining the low ejecta mass we observe for our sample. 




\subsection{Strongly stripped-envelope SNe}

As discussed in Section \ref{section:dis_lc}, the ejecta masses observed are consistent with theoretical predictions for stripped He stars with initial mass less than 3 \Msun. These occur in tight binaries where a helium star transfers mass to a small companion. The core-collapse of such stellar cores with a small envelope gives rise to low ejecta mass explosions. One way to form such progenitor stars in the stellar evolution theory is through the USSN scenario \citep{Tauris2013, Tauris2015}. These explosions are one of two channels to form double neutron star (DNS) binaries that are compact enough to merge within a Hubble time due to gravitational wave (GW) radiation.

\citet{Nakaoka2021} suggested that the Ca-rich IIb SN 2019ehk originated in an ultra-stripped core-collapse SN. Low-mass progenitors of stripped core-collapse SNe can retain a range of H and He masses depending on the nature of the companion and the initial binary period \citep{Yoon:2010aa, Zapartas2017, Laplace2020}. Presence of hydrogen represents a less extreme case of a `strongly stripped SN' (ejecta mass $\approx$ 0.6 \msun ) from a low mass (ZAMS $\approx$ 9 $-$ 12 \msun ) progenitor, that lost nearly all of its H envelope to stripping in a relatively compact orbit (orbital period $\sim$ 10 days). This is a transition class between the USSNe and the usual SESNe (see Figure \ref{fig:mnimej2}). The progenitor could be stripped by a main sequence companion and resulted in the formation of a compact neutron star $–$ main sequence binary. Such binaries have been suggested
to be progenitors of neutron star $-$ white dwarf systems that could merge within a Hubble time, and be detectable with LISA \citep{Toonen2018}. 

We compare our estimates of the O yields with that predicted for ultra-stripped SNe \citep{Tauris2013}. We use the nucleosynthetic yields from \citet{Moriya2017}, \citet{Yoshida2017} to estimate the [O~I] luminosity in the nebular phase for low-mass CO cores of 1.45 $–$ 1.5 \Msun, under different assumptions of the explosion energy and ejecta mass. Figure \ref{fig:olum} shows that the upper limits on the [O~I] luminosity for the low-mass CO core models are similar to the low [O~I] luminosity measured for our sample. Also, we find that the O masses estimated in the context of ultra-stripped SNe are consistent with the range estimated for our sample (see Section \ref{nebulara}), 


Ca-rich spectra strengthens the case for a strongly-stripped envelope scenario. Explosion of low-mass He-star progenitors can explain the dominance of Calcium emissions over the oxygen emissions (i.e., deficiency in the oxygen emissions) in the nebular spectra \citep{Kawabata2010, Maeda2007, Fang2019}. Also, the rapid evolution seen for our sample is a natural consequence from such a scenario. Such low-mass progenitor explosions are a boundary between the SN explosion (either by a Fe-core collapse or ONeMg-core electron capture) and WD formation, and predicts small ejecta mass, i.e., $<$ 0.5 \msun\ \citep{Tauris2013,Moriya2017}. These features are consistent with the observational properties of the Ca-rich Type IIb transients.

Thus, we conclude that the ejecta properties of our sample are generally consistent with the expectations for a low-mass ($<$ 2\ \msun) He (or C+O) star explosion corresponding to ZAMS mass of 8 $-$ 12 \msun , which defines a boundary between the CCSN explosion and a WD formation. Assuming a residual H mass of $\sim$ 0.1 \Msun, the inferred ejecta mass of our sample of $<$0.6 \Msun\  is consistent with a final CO core mass of $\approx$ 1.45 $-$ 1.5 \msun that collapses to form a $\approx 1.3$ \msun neutron star and ejects $\approx$ 0.5 \msun of material. The inferred ejecta mass points to a class of SSSNe that are potential progenitors of NS $-$ WD systems.

\subsection{Rate estimation}


\textcolor{black}{For this work, since all our sources were detected in the BTS survey, we follow the rate estimation analysis adopted by \citet{Perley2020}}. We discuss the procedure in brief below.

In the case of an ideal survey that is able to scan the entire sky to a given magnitude limit $m_{\rm lim}$, the volumetric rate can be estimated using

\begin{equation}
    R = \frac{1}{t} \sum_{i=1}^{N} \frac{1}{(\frac{4\pi}{3}D_{{\rm max},i}^3)},
\end{equation}

\noindent
where t is the duration of the survey, $D_{{\rm max},i}$ is the distance out to which the $i$th transient can be detected above $m_{\rm lim}$ at peak light in the absence of extinction, given its peak absolute magnitude $M_i$.
However, for practical surveys, there are multiple loss factors one needs to take into account.  For this work, we follow the revised equation adopted by \citet{Perley2020}:

\begin{equation}
    R = \frac{1}{t} \sum_{i=1}^{N} \frac{1}{(\frac{4\pi}{3}D_{{\rm max},i}^3)  f_{\rm sky} f_{\rm ext} f_{\rm rec}  f_{{\rm cl},i}}.
\end{equation}
\noindent
\textcolor{black}{The description of the loss factors and their values used are listed below.} 

\begin{itemize}

\item $f_{\rm sky}$ is the average active survey footprint expressed as a fraction of the full sky. \textcolor{black}{We use $f_{\rm sky} = 0.35$, based on the average active area of the BTS survey of $\sim 14400$ $\mathrm{deg^2}$.}

\item $f_{\rm ext}$ is the average reduction in effective survey volume owing to Galactic extinction. \textcolor{black}{We use $f_{\rm ext} = 0.82$, calculated by averaging the reduction in volume associated with the extinction toward each separate ZTF field. We do not correct for host-galaxy extinction.} 

\item $f_{\rm rec}$ is the average recovery efficiency for a detectable transient within the survey footprint: the probability that it is found and included in the sample.  \textcolor{black}{We use $f_{\rm rec} = 0.60$. This was measured by calculating the fraction of all events classified as SNe Ia that passed the quality cuts of the ZTF survey.}

\item $f_{{\rm cl},i}$ is the classification efficiency, which may depend on apparent magnitude. \textcolor{black}{We use $f_{\rm cl}=1.0$ if m $<$ 17.2 mag and $f_{\rm cl}=0.9$ at m $= 18.5$ mag, with a linear decline in between.}

\end{itemize}

Refer to \citet{Perley2020} for further details on the calculation of the loss parameters. The sample as presented in this paper spans $t = 3.5\ \mathrm{yr}$ of ZTF. We had around 94 sources classified as Type IIb as part of BTS and CLU surveys, out of which we have good quality nebular spectra for 29 Type IIb sources. \textcolor{black}{We use $H_0 = 70\ \mathrm{km} \mathrm{s}^{-1} \mathrm{Mpc}^{-1}$.} Based on these assumptions and the analysis adopted by \citet{Perley2020}, we get a rate of $(5.1^{+6.6}_{-4.5}) \times 10^3\  \textcolor{black}{\mathrm{Gpc}^{-3}}\mathrm{yr}^{-1} $. This is $\approx 6^{+12}_{-5}\% $ of CCSNe and $\approx 22^{+25}_{-19}\%$ of SNIa assuming a CCSNe rate of $(7.7^{+6.6}_{-4.5})\times10^{4}\ \mathrm{Gpc}^{-3} \mathrm{yr}^{-1}$ and SNIa rate of $(2.35 \pm 0.24)\times10^{4}\ \mathrm{Gpc}^{-3} \mathrm{yr}^{-1}$ calculated using the same procedure. As per present classifications, this is $\approx 22 \pm 6\%$ of all SESNe and $\approx 44 \pm 31\%$ of Type IIb SNe. We note that the rates are based on the present classification status. Also, since these events have low luminosity and are low in number, the associated errors are high. We leave a more accurate rate estimate using a larger sample size and more complete nebular follow-up for future work.

NS-WD merger rates are predicted to be in the range of 3-15\% of Type Ia supernova rate \citep{Toonen2018}. \textcolor{black}{The corresponding rates are consistent with that obtained for the Ca-rich Type IIb sample within the large errors, at the low end of the relative rate estimate.} Also, the delay time distribution suggests that these mergers are most likely to be found in star-forming galaxies, consistent with the host galaxy properties of our sample.




\section{Conclusion} \label{sec:conclusion}

We examined the various photometric and spectroscopic properties of the objects in our sample of Ca-rich Type IIb SNe. We find that the observed properties of our sample suggest a class of Strongly-stripped SNe (SSSNe) that form a transition class between the usual Stripped-Envelope SNe (SESNe) and Ultra-Stripped SNe (USSNe) groups. The key takeaways of the paper are:

\begin{enumerate}
    
    \item 
    SSSNe are a class of core-collapse SNe which have ejecta mass less than $\sim$ 1 \Msun\ but show hydrogen in the ejecta. They have been classified as Type IIb SNe based on the photospheric spectra. The SNe in our sample also shows a high [Ca~II]/[O~I] flux ratio ($\geq 2$) in the nebular spectra. Observed properties suggest that the progenitor could be stripped by a main sequence companion and may result in the formation of a compact neutron star $–$ main sequence binary. Such binaries have been suggested to be progenitors of neutron star $-$ white dwarf systems that could merge within a Hubble time, and be detectable with LISA.
    
    \item
    The ejecta and spectroscopic properties of the class are generally consistent with the expectations for a low-mass ($<$ 3\msun) He star explosion corresponding to a ZAMS mass less than 12 \msun. From the low [O I] $\lambda \lambda$6300, 6364  luminosity in the nebular phase spectra, we find that the estimated progenitor oxygen mass \textcolor{black}{lower limits} for our sample is in the range \textcolor{black}{$\approx$ $0.005-0.1$ \msun}. For core-collapse SNe models to reproduce our inferred oxygen mass, we infer that the ZAMS mass is less than 12 \msun. Other measured properties such as the [Ca~II]/[O~I] ratio, ejecta mass, kinetic energy, envelope radius are also consistent with the 8 $-$ 12 \Msun\ progenitor mass scenario.

    \item 
    
    The presence of photospheric hydrogen suggests $M_{H} \gtrsim 0.02-0.03\ M_\odot$ \citep{Hachinger2012}. The proposed explosion mechanism for the usual Ca-rich Type I  events involves CO + He binary WDs progenitors. However they are expected to be very deficient in hydrogen ($M_{H}$ $ < 10^{-4} M_\odot$). Also, the host galaxies of the SSSN class are consistent with star-forming galaxies as they lie in the galaxy main sequence of the mass-SFR plane and have a luminosity range consistent with the hosts for the general Type IIb SN population. Hence, our sample is distinct from the thermonuclear Ca-rich gap transients found in old environments.

\end{enumerate}

Thus, we present a class of SNe that lie in the low mass end of CCSNe and are a transition class between USSNe and the usual SESNe. Observational constraints for such low mass He stars are missing in the literature. This sample provides an opportunity to understand the low mass He star evolution, the extent of binary stripping, and would help in developing theoretical models including nebular spectra for this low mass range. Discovery and follow-up of more members of this class will help obtain a stricter constraint on the rates of these events.




    
    

\section{Data availability}

All the photometric and spectroscopic data used in this work can be found  
 \href{https://caltech.box.com/s/q4sge6ujnzd86xeuodwqekovq8y69lcd}{here}.

The optical photometry and spectroscopy will also be made public through WISeREP, the Weizmann Interactive Supernova Data Repository \citep{Yaron2012}.


\acknowledgments
We thank Adam Burrows for valuable comments on the manuscript. We thank the anonymous referee for their valuable comments that significantly improved the quality of the paper.
Based on observations obtained with the Samuel Oschin Telescope
48-inch and the 60-inch Telescope at the Palomar Observatory as part of the Zwicky Transient Facility project.
ZTF is supported by the National Science Foundation under Grant No. AST-2034437 and a collaboration including
Caltech, IPAC, the Weizmann Institute of Science, the Oskar Klein Center at Stockholm University, the University
of Maryland, Deutsches Elektronen-Synchrotron and Humboldt University, the TANGO Consortium of Taiwan, the
University of Wisconsin at Milwaukee, Trinity College Dublin, Lawrence Livermore National Laboratories, IN2P3,
France, the University of Warwick, the University of Bochum, and Northwestern University. Operations are conducted
by COO, IPAC, and UW. 

SED Machine is based upon work supported by the National Science Foundation under
Grant No. 1106171. 

The ZTF forced-photometry service was funded under the Heising-Simons Foundation grant \#12540303 (PI: Graham).

The GROWTH Marshal was supported by the GROWTH project funded by the National Science Foundation under Grant No 1545949.

The data presented here were obtained in part with ALFOSC, which is provided by the Instituto de Astrofisica de Andalucia (IAA) under a joint agreement with the University of Copenhagen and NOT.

The Liverpool Telescope is operated on the island of La Palma by Liverpool John Moores University in the Spanish Observatorio del Roque de los Muchachos of the Instituto de Astrofisica de Canarias with financial support from the UK Science and Technology Facilities Council.
Based on observations made with the Italian Telescopio Nazionale Galileo (TNG) operated on the island of La Palma by the Fundación Galileo Galilei of the INAF (Istituto Nazionale di Astrofisica) at the Spanish Observatorio del Roque de los Muchachos of the Instituto de Astrofisica de Canarias.

The W. M. Keck Observatory is operated as a scientific partnership among the California Institute of Technology, the University of California and the National Aeronautics and Space Administration. The Observatory was made possible by the generous financial support of the W. M. Keck Foundation. The authors wish to recognize and acknowledge the very significant cultural role and reverence that the summit of Maunakea has always had within the indigenous Hawaiian community.  We are most fortunate to have the opportunity to conduct observations from this mountain.
The \texttt{ztfquery} code was funded by the European Research Council (ERC) under the European Union's Horizon 2020 research and innovation programme (grant agreement $n^{\circ}759194 -$ USNAC, PI: Rigault).

S. Schulze acknowledges support from the G.R.E.A.T research environment, funded by {\em Vetenskapsr\aa det},  the Swedish Research Council, project number 2016-06012.

\FloatBarrier

\bibliography{main}
\bibliographystyle{aasjournal}

\appendix

\begin{figure}
\centering
\includegraphics[width=1\columnwidth]{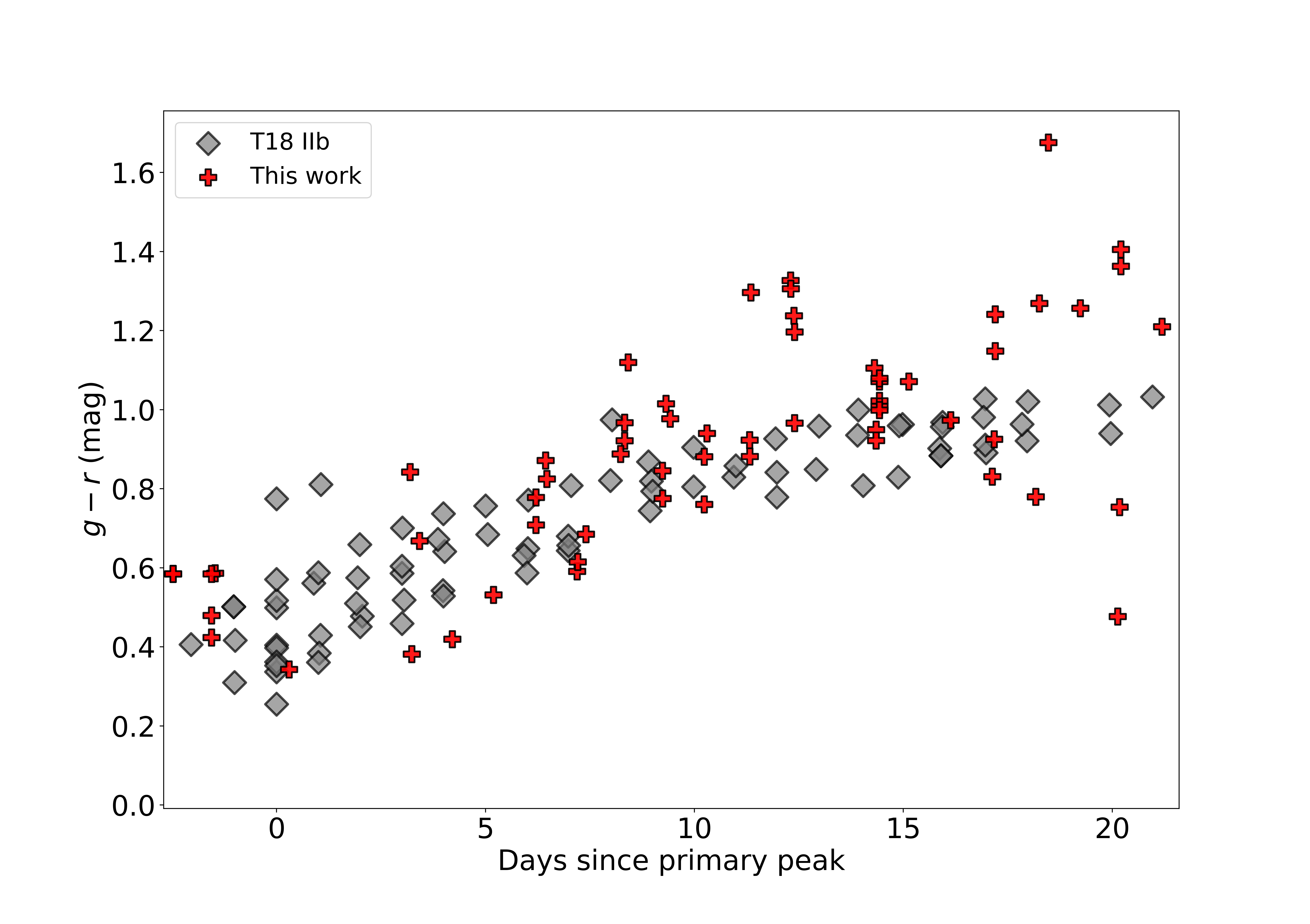}
\caption{The $g-r$ color evolution of the SNe in our sample represented by red crosses compared to the Type IIb SNe sample of \citet{Taddia2018} in gray diamonds.}
\label{fig:color}
\end{figure}


\section{Nebular Spectra} \label{appendix:nebfull}
\begin{figure*}[!h]
    \centering
    \includegraphics[width=18cm]{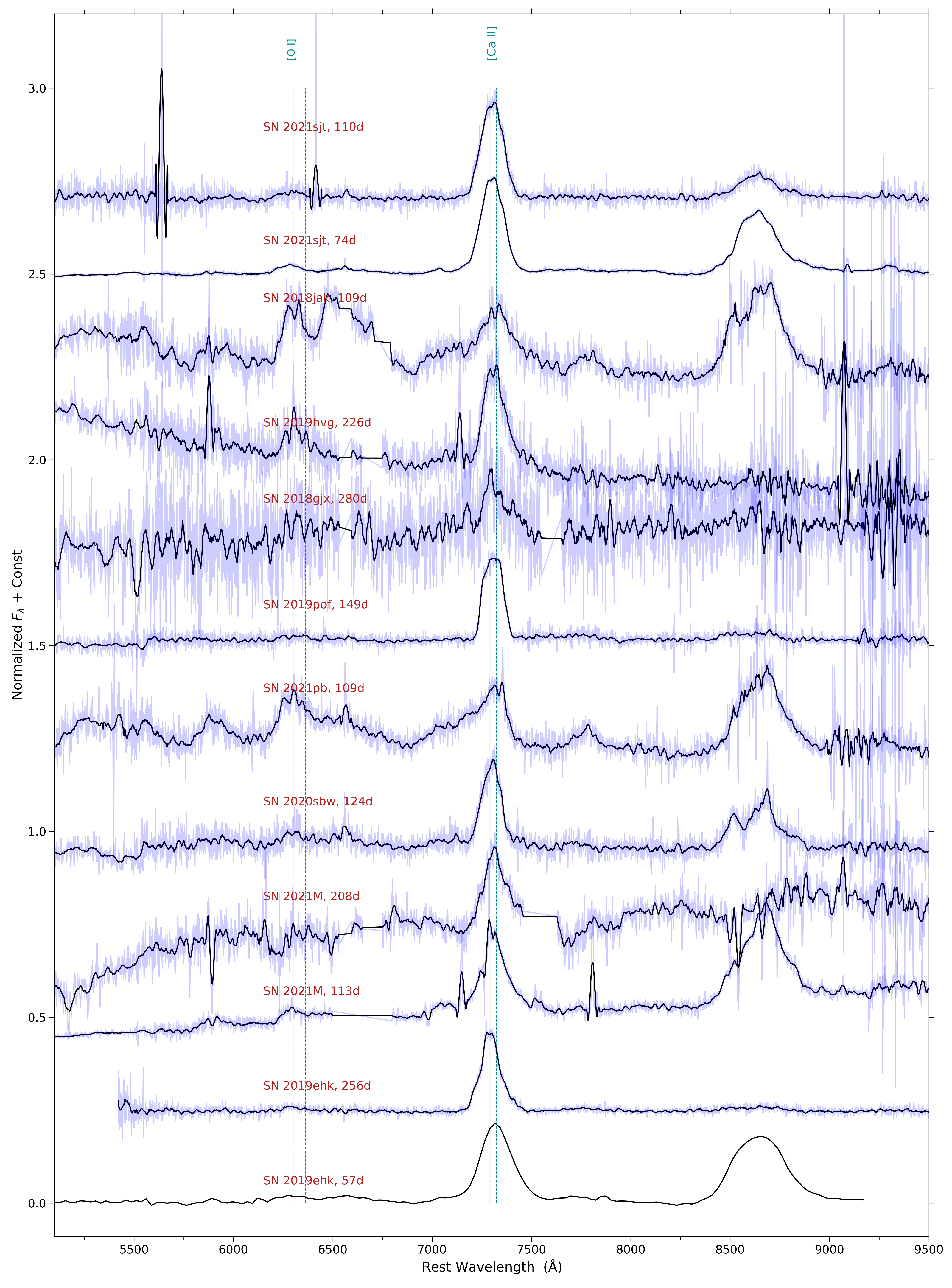}
    \caption{\textcolor
    {black}{The full-wavelength range of the nebular spectra shown in Figure \ref{fig:all_nebspectra}. The phase of the spectra is measured as the number of rest-frame days since the second peak.}}
    \label{fig:fullneb}
\end{figure*}

The full-wavelength range nebular spectra are shown in Figure \ref{fig:fullneb}. The phase of the spectra is measured as the number of rest-frame days since the second peak. 

\section{Lightcurve Modeling fits} \label{appendix:piro}
\begin{figure*}[!h]
    \centering
    \includegraphics[width=18cm]{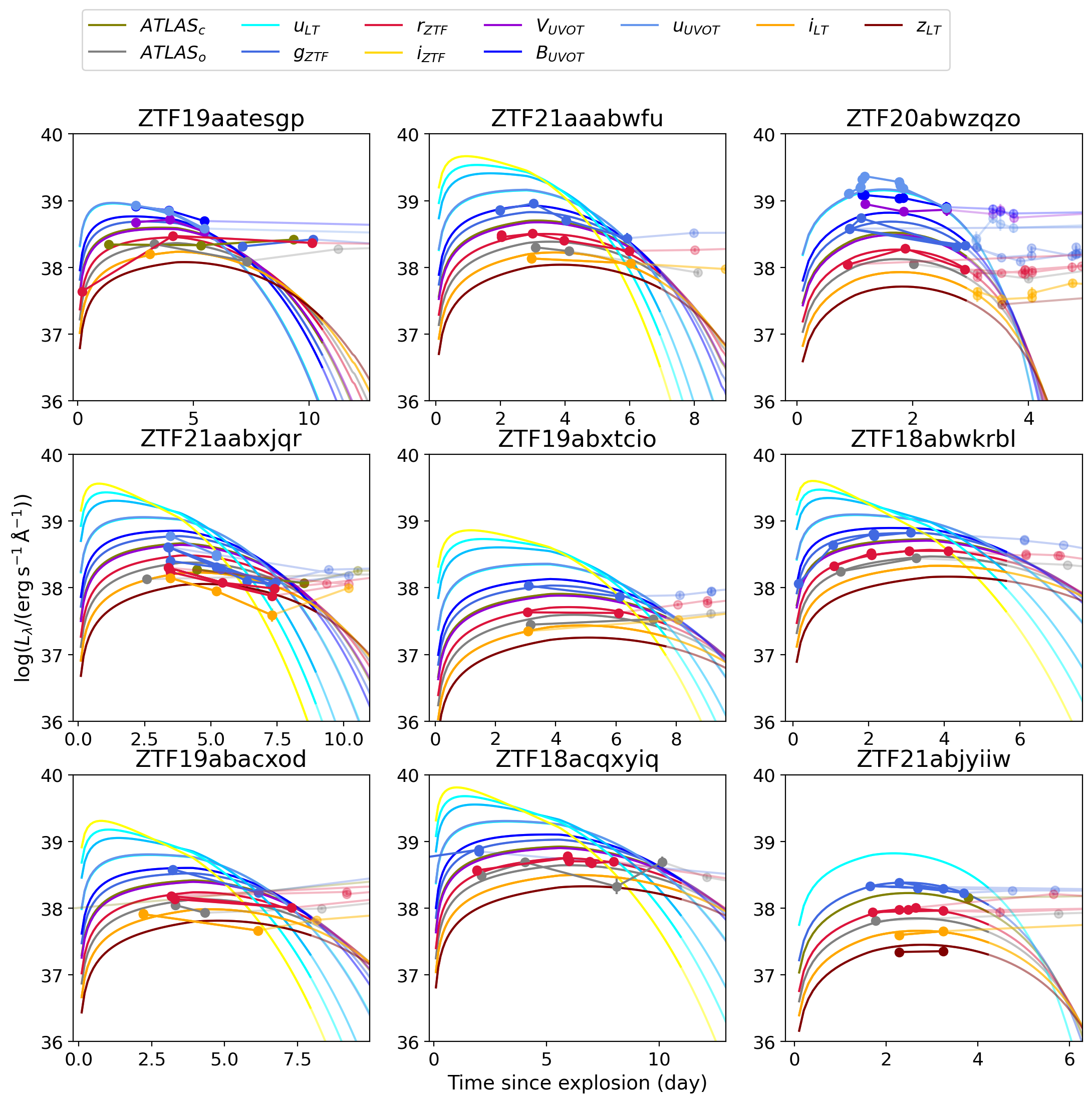}
    \caption{Shock cooling model \citep{Piro2021} fits to multi-band data.}
    \label{fig:sb_collage}
\end{figure*}

\begin{figure*}[!h]
    \centering
    \includegraphics[width=18cm]{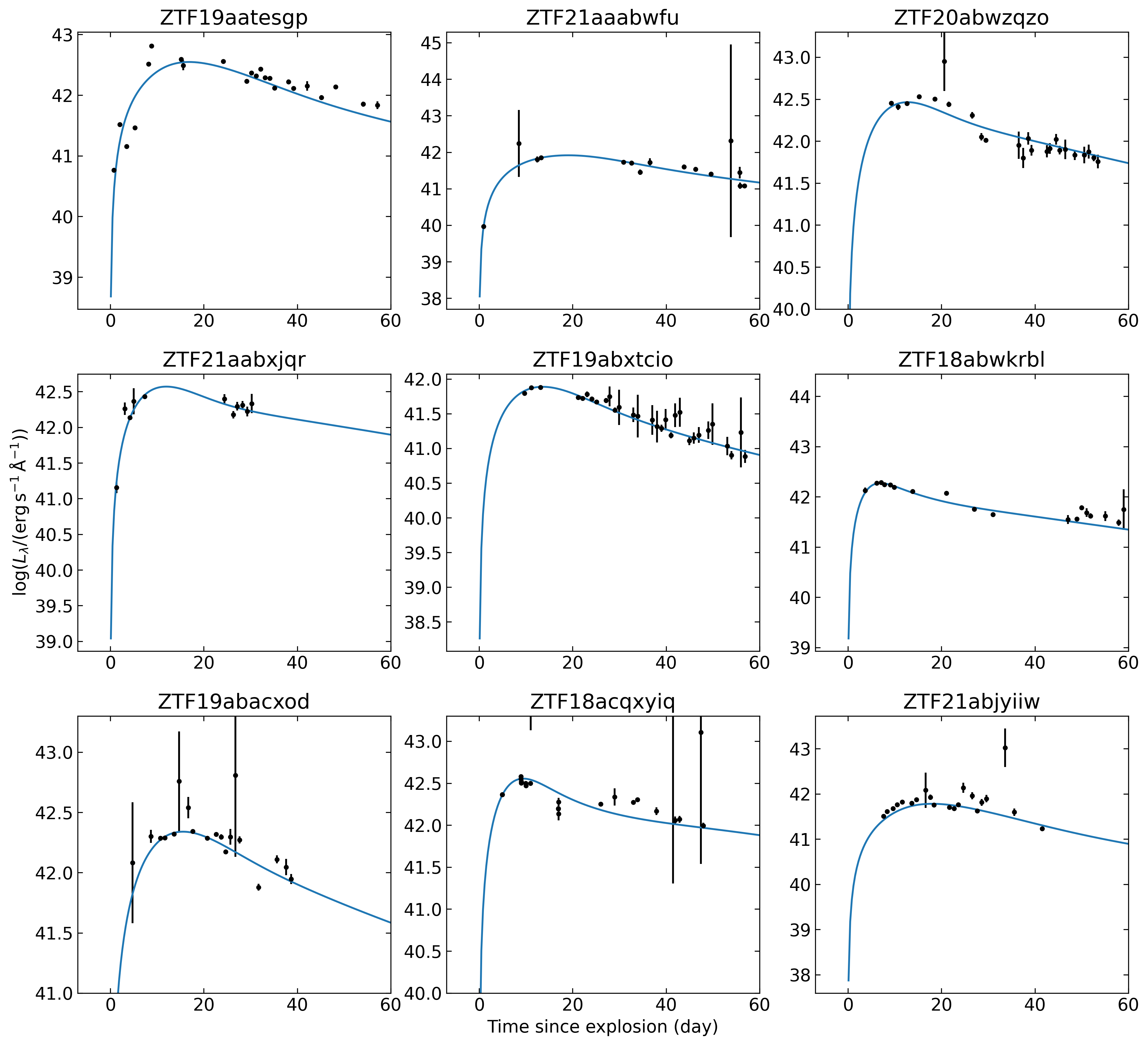}
    \caption{Radioactive \citep{Arnett89} fits to bolometric luminosity data excluding the data points fit by shock-cooling.}
    \label{fig:arnett_collage}
\end{figure*}

Figure \ref{fig:sb_collage} shows a collage of the best-fit lightcurves for the shock cooling model \citep{Piro2021} fits to the multi-band photometry data. Figure \ref{fig:arnett_collage} shows a collage of the best-fit lightcurves for the radioactive peak model \citep{Arnett89} fits to the bolometric luminosity data.

\section{Photometry and spectroscopy data}
All the photometry and spectroscopy data are provided as machine-readable tables  \href{https://caltech.box.com/s/q4sge6ujnzd86xeuodwqekovq8y69lcd}{here.}

\section{Blackbody fits} \label{appendix:bbfit}
All the best-fit parameters including bolometric luminosity, radius and temperature for each object are provided as machine-readable tables  \href{https://caltech.box.com/s/q4sge6ujnzd86xeuodwqekovq8y69lcd}{here.}

\begin{table*} 
\begin{center} 
\caption{Summary of the blackbody properties for SN 2021sjt (Truncated). All the best-fit parameters including bolometric luminosity, radius and temperature for each object will be provided as machine-readable tables after publication \href{https://caltech.box.com/s/q4sge6ujnzd86xeuodwqekovq8y69lcd}{here.} } 
\begin{tabular}{cccc} 
\hline  \\  

Phase &  Log Luminosity  &  Temperature & Radius   \\    (days since first detection)   & ($\mathrm{erg\ s^{-1}}$)  & (K) & (\Rsun) \\\\

\hline  \\
$2$  &  $42.34^{+0.56}_{-0.21}$  &  $14615^{+12882}_{-4534}$  &  $3728^{+2424}_{-1711}$ \\ \hline \\ 

$3$  &  $42.10^{+0.02}_{-0.02}$  &  $8449^{+520}_{-438}$  &  $8445^{+786}_{-774}$ \\ \hline \\ 

$5$  &  $42.08^{+0.01}_{-0.01}$  &  $7118^{+340}_{-321}$  &  $11716^{+1013}_{-912}$ \\ \hline \\ 

$6$  &  $42.07^{+0.01}_{-0.01}$  &  $6580^{+252}_{-242}$  &  $13412^{+1001}_{-912}$ \\ \hline \\

\end{tabular}  \label{bb_table} 
\end{center} 
\end{table*}

\begin{table*} 
\begin{center} 
\caption{Summary of the photometry data used for SN 2021sjt (Truncated). The photometry and spectroscopy data for each object will be provided as machine-readable tables after publication  \href{https://caltech.box.com/s/q4sge6ujnzd86xeuodwqekovq8y69lcd}{here.} } 
\begin{tabular}{ccccc} 
\hline  \\  

Source &  date  &  filter & mag & limiting mag   \\   & (JD)  & (AB mag) & (AB mag) & (AB mag) \\\\ \hline  \\

ZTF21abjyiiw  &  $2459397.43$  &  ztfg  &  $nan $  &  $20.28$ \\ \hline \\ 

ZTF21abjyiiw  &  $2459400.39$  &  ztfg  &  $nan $  &  $20.58$ \\ \hline \\ 

ZTF21abjyiiw  &  $2459400.43$  &  ztfr  &  $nan $  &  $20.18$ \\ \hline \\ 

ZTF21abjyiiw  &  $2459402.35$  &  ztfg  &  $18.58 \pm 0.08$  &  $19.74$ \\ \hline \\ 

ZTF21abjyiiw  &  $2459402.35$  &  ztfg  &  $18.58 \pm 0.08$  &  $19.74$ \\ \hline \\ 

ZTF21abjyiiw  &  $2459402.40$  &  ztfr  &  $18.01 \pm 0.06$  &  $19.36$ \\ \hline \\

\end{tabular}  \label{phot_table} 
\end{center} 
\end{table*}
\section{Host galaxy modeling} \label{appendix:host}
The models used for \cigale{} and the summary of the host-galaxy best-fit parameters are provided in Table \ref{tab:cigalemod} and Table \ref{tab:galaxyprop} respectively.
\begin{table*}
\begin{center}
\caption{The models and fitting parameters used for \cigale{} }

\begin{tabular}{ cc }

\hline 
Galaxy attributes & Brief description \\\\
\hline 
SFH & SFR $\propto$ $\frac{t}{\tau^2}$ $exp(-t/\tau)$ \\
& $\tau= 250, 500, 1000, 2000, 4000, 6000, 8000$~Myr  \\
& Age $t = 250, 500, 1000, 2000, 4000, 8000,  12000$~Myr \\
\hline \\
SSPs models & BC03 \citep{Bruzual2003} $+$ Chabrier IMF \citep{Chabrier2003} \\
& Stellar metallicity = 0.0004, 0.004, 0.008, 0.02 $Z_\odot$  \\
\hline \\
Dust Attenuation & Modified power law curves \citep {Calzetti2000} + differential reddening of stars according to age \\
& $E(B-V)_{\rm young}$ = 0.0, 0.2, 0.3, 0.5, 0.8,  1.0, 1.5, 2.0, 3.0 \\
& $E(B-V)_{\rm old}$ = 0.3, 0.50, 1.0 \\
& ${\rm UV-bump~wavelength} = 217.5~{\rm nm}$\\
& ${\rm UV-bump~amplitude} = 0.0, 1.0, 2.0, 3.0 $\\
& ${\rm powerlaw~slope} = -0.13, -0.2 , -0.5$ \\
\hline \\
Dust emission &  Dust templates of \citet{Dale2014} $+$ Energy Balance  \\
& AGN fraction = 0 \\
& alpha = 1.0, 1.5, 2.0, 2.5 \\
\hline \\
Nebular & $\log U_{\rm ion} =$ -2.0 \\
& emission line width = 300.0 km/s  \\
\hline
\end{tabular} \label{tab:cigalemod}
\end{center}
\end{table*}

\begin{table*} 
\begin{center} 
\caption{Summary of the host-galaxy best-fit parameters}

\begin{tabular}{cccc} 
\hline  \\  

ZTF Name & log Mass & log SFR & B-band abs. mag   \\ &  (\Msun) & (\Msun/yr) & (mag)  \\\\ \hline  \\ 

SN 2019ehk/ZTF19aatesgp  &  $11.09^{+0.01}_{-0.00}$  &  $0.23^{+0.01}_{-0.00}$  &  $\textcolor{black}{-21.87}$    \\ \hline \\ 

SN 2021M/ZTF21aaabwfu  &  $9.88^{+0.11}_{-0.15}$  &  $0.21^{+0.29}_{-0.12}$  &  $-18.75$     \\ \hline \\ 

SN 2020sbw/ZTF20abwzqzo  &  $10.71^{+0.06}_{-0.01}$  &  $-0.17^{+0.07}_{-0.01}$  &  $-20.14$    \\ \hline \\ 

SN 2021pb/ZTF21aabxjqr  &  $10.74^{+0.20}_{-0.04}$  &  $0.67^{+0.16}_{-0.29}$  &  $-21.17$     \\ \hline \\ 

SN 2019pof/ZTF19abxtcio  &  $8.62^{+0.01}_{-0.01}$  &  $-2.32^{+0.09}_{-0.07}$  &  $-15.67$     \\ \hline \\ 

SN 2018gjx/ZTF18abwkrbl  &  $10.25^{+0.11}_{-0.11}$  &  $0.42^{+0.21}_{-0.12}$  &  $-19.84$     \\ \hline \\ 

SN 2019hvg/ZTF19abacxod  &  $9.93^{+0.11}_{-0.13}$  &  $0.55^{+0.09}_{-0.17}$  &  $-20.15$   \\ \hline \\ 

SN 2018jak/ZTF18acqxyiq  &  $9.84^{+0.26}_{-0.36}$  &  $1.09^{+0.55}_{-0.23}$  &  $-19.52$    \\ \hline \\ 

SN 2021sjt/ZTF21abjyiiw  &  $10.70^{+0.18}_{-0.39}$  &  $2.11^{+1.44}_{-0.99}$  &  $-16.78$    \\ \hline \\ 

\end{tabular}  \label{tab:galaxyprop} 
\end{center} 
\end{table*}

\end{document}